\documentclass{aa}
\usepackage[dvips]{graphics}
\usepackage{latexsym}

\begin{document}

\thesaurus{3(11.19.6;11.06.2;11.04.1;13.09.1)}

\title{Near-infrared observations of galaxies in Pisces-Perseus:
III. Global scaling relations of disks and bulges}

\author{ G. Moriondo \inst{1}
\and
R. Giovanelli \inst{2}
\and
M.P. Haynes \inst{2} 
}
\institute{ 
  Dipartimento di Astronomia e Scienza dello spazio, Universita' di Firenze,
L. E. Fermi 5, I-50125 Firenze, Italy
 \and  
  Center for Radiophysics and Space Research, Cornell University, Ithaca, NY 
14853 
%%  \and 
%%   ??
} 

\offprints{gmorio@arcetri.astro.it}
\date{Received ; accepted }

\maketitle
\markboth{Moriondo et al.: Spiral scaling relations}{Moriondo et al.: Spiral scaling relations}

\begin{abstract}
We determine the parameters of scaling relations analogous to the 
Fundamental Plane of 
elliptical galaxies for the bulges and disks of a sample of 40 spiral galaxies. 
To this end we derive structural parameters (scalelengths and 
surface brightnesses) from near infrared $H$ band images, and kinematical 
parameters (rotational velocities) from optical rotation curves.
In the case of the disks, we test the accuracy of the derived relation as a 
distance indicator by comparing its scatter to that of the $H$ band 
Tully--Fisher relation for the same sample, and find 
that the accuracy attained by the latter is slightly 
higher (the dispersion is 19~\% versus 23~\% for this sample).
It is speculated that the difference is due to the more robust character of
global parameters, rather than those associated with the inner parts of
disks. It also apperas that (a) either the stellar mass-to-light ratios of 
bulge and disk increase with the size of the components, or (b) the bulge 
and disk relative contributions to the overall rotation of the galaxy (and, as 
a consequence, to its total mass) become steadily smaller with increasing size.
\end{abstract}

\section{Introduction}

The existence of tight scaling relations between observable photometric
and kinematic galaxy parameters, in particular the Fundamental Plane (FP) of 
elliptical galaxies (J\o rgensen et al. \cite{jorg}, Scodeggio
\cite{scodeggio}), and the Tully-Fisher relation for spirals (TF, 
Tully \& Fisher \cite{tf}), finds 
its most straightforward application in the evaluation of galaxy distances. 
The small scatter observed in both relations implies a fine tuning 
between parameters strictly related to the stellar component alone, such 
as the optical luminosity, 
and the kinematic properties of the galaxy, which are affected 
by the overall mass distribution (both dark and visible). 

% In general, for a self gravitating system we expect a connection, just on the
% basis of a dimensional analysis of the virial theorem, between any
% characteristic velocity, mass surface density, and scalelength $r$ of the
% system:
% \begin{equation}
% \protect\label{eq:virt}
% r \sim v^2 \, \Sigma^{-1}
% \end{equation}
% The proportionality constant depends on the geometry of the system
% and on the choice of the quantities that appear in the equation
% (for example, for a rotating, exponential, thin disk we can choose the
% exponential folding length, the central surface brightness and the peak
% rotation velocity at 2.2 $R_d$).
% If the mass-to-light ratio is constant throughout the whole
% system the same relation also holds when the surface brightness is used
% instead of $\Sigma$.
% In this hypothesis, if we consider a sample of objects with the same shape,
% and the same mass-to-light ratio, we expect Eq. (\ref{eq:virt}) to hold,
% with the same proportionality constant, provided that
% the same parameters are measured for every object. 
% Both TF relation and FP deviate from Eq. (\ref{eq:virt}), and both exhibit a 
% small scatter, implying that all {\it all} the parameters that define
% an elliptical or a spiral galaxy from the photometric and dynamical 
% point of view are tightly correlated with the quantities in Eq. (\ref{virt}).
% The mentioned scaling relations, therefore, need to find their origin 
% end explanation in the process of galaxy formation. 

In the case of spiral galaxies, it is therefore suggestive that a tight 
scaling relation may also exist between photometric and kinematic properties 
{\it of the disks alone}, with a scatter as low as, or even lower, than the 
TF relation. 
Karachentsev (\cite{kara}) found that, for a sample of 15 edge-on galaxy, 
the disk scale length $R_d$ was connected to the 21~cm line velocity width
$W$ and the $I$--band central disk brightness $I(0)$ by the relation 
$R_d \propto W^{1.4} I(0)^{-0.74}$, 
with a scatter in $\log (R_d)$ of 0.048. The uncertainty
implied in the distance is around 12\%, smaller than the one usually 
achieved by the TF relation ($15 \sim 20$\%). 
More recently Chiba \& Yoshii (\cite{cy}, CY95 hereafter) tested on a sample 
of 14 nearby spirals the relation 
\begin{equation}
\protect\label{eq:cy}
\log R_d = a \log (V_2\; I(0)^{-0.5}) + b
\end{equation} 
in the $B$ band,
where $V_2$ is the galaxy rotation velocity measured at 2.2 disk scalelengths.
Since the contribution from the disk to the overall rotation curve (RC) 
has a maximum at this particular distance, the authors 
argue that the contributions from bulge and dark 
halo are likely to be less important, and therefore the 
measured velocity should represent a good estimate for the rotation of the 
disk alone.
For a set of exponential disks of fixed mass--to--light ratio ($M/L$) one 
would expect $\log R_d \sim \log \, [V_2^2/I(0)]$, corresponding to $a=2$ 
in Eq. 
(\ref{eq:cy}). Actually CY95 find $a=1.045$, again with a remarkably small 
scatter. 
On the other hand, recent work from Giovanelli (\cite{giovatf}) suggests 
that the accuracy of Eq. (\ref{eq:cy}) as a distance indicator 
is inferior to the one attained by the traditional TF relation 
by at least a factor of 2.

Besides beeing a tool to provide redshift-independent distances,
scaling relations also contain information about 
how galaxies -- and in particular their ``visible'' constituents -- have formed 
and evolved (Gavazzi \cite{gava}, Ciotti \cite{ciotti}, Dalcanton et al.
\cite{dalca}, Burstein et al. \cite{burst} -- BBFN hereafter). 
In this respect, spiral disks are a potentially ``easy'' class of systems,
since they are all characterized by well defined shape and 
kinematics: if the effects of extinction on their surface 
brightness distribution are accounted for, and a reliable estimate of their 
mass is obtained, the subsequent scaling relations will be directly related 
to systematic variations in the disks' stellar content.
Further information can be provided by the variation with wavelength 
of such properties, in particular by scaling relations involving 
colours (e.g., the colour magnitude relation; see Gavazzi \cite{gava1},
Tully et al. \cite{tully}, Peletier \& de Grijs \cite{pdg}). 

In this work we investigate the existence and tightness of a general scaling
relation in the near infrared (NIR), similar to the one defining the FP, for 
both structural components (disks and bulges) of a sample of 40 nearby spiral 
galaxies.
Our aim is to test the power of these relations as tools to measure galaxy 
distances, and use them to provide new information about the stellar content 
and star formation history of spiral galaxies.
In the present paper we are going to deal mainly with the problem of the
distance measurements, leaving a thorough discussion of the second point 
to future work.
The use of NIR photometry is well suited for a study of this kind,
since it minimizes the effect of internal extinction, and 
provides a good tracing of the stellar mass.
In most cases, high resolution, optical RC's 
allow us to trace the gravitational potential of the galaxies up to their 
innermost regions.

The photometric parameters are obtained, in the NIR $H$ band, from a
bi-dimensional decomposition of the galaxy images, as described in 
Sect.~\ref{phot}.
In Sect. \ref{kin} we show how the kinematical information are extracted 
from the galaxy RC's, to which we fit a model composed by 
bulge, disk, and a dark halo. 
We subsequently derive the coefficients of the FP for bulges 
and disks, and compare the potential accuracy in a distance determination 
achieved by the disks' relation to the one we obtain using the TF relation 
for the same sample.
We finally investigate the trends of $M/L$'s with luminosity and galaxy size.
Throughout the paper we adopt a Hubble constant 
$H_{\circ} = 75$ km s$^{-1}$ Mpc$^{-1}$.

\section{The data}

The 40 spiral galaxies considered for this study are drawn from a larger 
sample selected in the Pisces--Perseus supercluster region, for which 
$H$ band images are available (see Moriondo et al. \cite{noi_2} and 
\cite{noi_1} for a thorough description of the original sample). 
% We use such images
% to derive the photometric and structural properties of the galaxies, 
% as described in the next section.
This subset in particular contains the galaxies for 
which a RC is also available both from the literature or the 
private database of RG and MH, and includes morphological types ranging 
from Sa to Scd. 
In most cases, the RC's are derived from optical emission lines measurements
and are confined within two or three disk scale lengths.
For a few galaxies radio aperture--synthesis RC's (21~cm HI line)
are also available, extending well beyond the optical radius. 
All the RC's were rescaled to our adopted values of distance 
and inclination.
For all the galaxies (except UGC~2885) 21~cm line velocity widths are also 
available from the database of RG and MH. We will use these values later on 
to derive the $H$ band TF relation for the sample.
Table \ref{table:sample} contains the basic information on the galaxies 
of the sample as well as the references for the RC's; many of these were
retrieved from the compilations by Corradi \& Capaccioli (\cite{corra})
and Prugniel et al. (\cite{prugni}).
We note that this sample is not appropriate to obtain an absolute
calibration of distances. It is however well suited to compare two
different scaling relations and their relative accuracy as distance
indicators. 

\begin{table*}
\caption{The galaxy sample}
\begin{flushleft}
\protect\label{table:sample}
\begin{tabular}{lllllllll}
\noalign{\smallskip}
\hline
\noalign{\smallskip}
\multicolumn{2}{c}{Names} & \multicolumn{2}{c}{R.A.~~~(J2000)~~~Dec.} & \multicolumn{1}{c}{RH Type} &
\multicolumn{1}{c}{$m_B$} & \multicolumn{1}{c}{$P.A.$} & \multicolumn{1}{c}{$D$ (Mpc)} & 
\multicolumn{1}{c}{RC code} \\
\multicolumn{1}{c}{(1)} & \multicolumn{1}{c}{(2)} & \multicolumn{1}{c}{(3)} &
\multicolumn{1}{c}{(4)} & \multicolumn{1}{c}{(5)} & \multicolumn{1}{c}{(6)} &
\multicolumn{1}{c}{(7)} & \multicolumn{1}{c}{(8)} & \multicolumn{1}{c}{(9)} \\
\noalign{\smallskip}
\hline
\noalign{\smallskip}
UGC~00014   &          & 00 03 35.0 & 23 12 03 & Sc          & 13.8 &  41 & 92 & 1,2 \\
UGC~00026   & NGC~7819 & 00 04 24.5 & 31 28 19 & SB(s)b      & 14.2 &  99 & 61 & 3 \\
UGC~00089   & NGC~0023 & 00 09 53.4 & 25 55 25 & SB(s)a      & 12.8 & 171 & 56 & 4 \\
UGC~00673   &          & 01 06 09.7 & 31 24 24 & SAc?        & 15.7 &  45:& 79 & 1 \\
UGC~00725   &          & 01 10 11.0 & 43 06 35 & SBcd?       & 14.4 &  47 & 63 & 5 \\
UGC~00732   &          & 01 10 44.4 & 33 33 28 & SA(r)d      & 14.6 &  81 & 68 & 1 \\
UGC~00820   & NGC~0452 & 01 16 14.8 & 31 02 01 & SBab        & 13.6 &  35 & 62 & 6 \\
UGC~00927   & NGC~0496 & 01 23 11.5 & 33 31 46 & Sbc         & 14.1 &  38 & 76 & 1,7 \\
UGC~00940   &          & 01 23 37.9 & 34 34 11 & SA(s)c      & 15.1 &  68 & 89 & 7 \\
UGC~00975   &          & 01 25 15.6 & 34 21 34 & S           & 15.0 & 123 & 61 & 7 \\
UGC~01013   & NGC~0536 & 01 26 21.6 & 34 42 14 & SB(r)b      & 13.2 &  72 & 65 & 7 \\
UGC~01033   &          & 01 27 34.9 & 31 33 16 & Scd:        & 14.1 & 132 & 50 & 7 \\
UGC~01094   & NGC~0582 & 01 31 57.7 & 33 28 33 & SB?         & 14.1 &  60 & 54 & 7 \\
UGC~01238   & NGC~0668 & 01 46 22.6 & 36 27 39 & Sb          & 13.7 &  30 & 56 & 7,8 \\
UGC~01302   & NGC~0688 & 01 50 44.1 & 35 17 04 & (R')SAB(rs)b& 13.3 & 141 & 52 & 7,8 \\
UGC~01350   &          & 01 52 57.5 & 36 30 47 & SB(r)b      & 14.2 &  44 & 67 & 7 \\
UGC~01437   & NGC~0753 & 01 57 42.4 & 35 54 57 & SAB(rs)bc   & 13.0 & 131 & 62 & 7,8,9,10 \\
UGC~01633   & NGC~0818 & 02 08 44.4 & 38 46 36 & SABc:       & 13.2 & 114 & 53 & 9,11 \\
UGC~01835   & IC~0221  & 02 22 41.3 & 28 15 28 & Sc          & 13.7 &   5:& 65 & 1 \\
UGC~01935   & NGC~0931 & 02 28 14.5 & 31 18 42 & Sbc         & 14.5 &  82 & 64 & 1,12 \\
UGC~01937   & NGC~0935 & 02 28 10.9 & 19 35 59 & Scd:        & 13.6 & 153 & 52 & 13 \\
UGC~02134   &          & 02 38 51.8 & 27 50 50 & Sb          & 14.2 & 100 & 58 & 1 \\
UGC~02142   & NGC~1024 & 02 39 12.2 & 10 50 50 & (R')SA(r)ab & 13.1 & 158 & 44 & 14 \\
UGC~02185   &          & 02 43 11.4 & 40 25 34 & Scd:        & 13.5 & 141 & 55 & 1 \\
UGC~02223   &          & 02 45 14.4 & 35 11 21 & Scd:        & 14.8 &  46 & 63 & 1 \\
UGC~02241   & NGC~1085 & 02 46 25.3 & 03 36 26 & SA(s)bc:    & 13.1 &   2 & 88 & 15 \\
UGC~02617   &          & 03 16 00.7 & 40 53 08 & SAB(s)d     & 13.8 &  10 & 59 & 7 \\
UGC~02655   &          & 03 18 45.3 & 43 14 20 & SAB(s)d     & 13.5 & --  & 80 & 7 \\
UGC~02885   &          & 03 53 02.4 & 35 35 22 & SA(rs)c     & 13.5 &  45 & 76 & 10,16 \\
UGC~11973   &          & 22 16 50.4 & 41 30 13 & SAB(s)bc    & 12.9 &  42 & 52 & 17 \\
UGC~12173   &          & 22 43 52.0 & 38 22 37 & SAB(rs)c    & 13.5 & 101 & 59 & 1,7 \\
UGC~12230   & NGC~7407 & 22 53 21.2 & 32 07 44 & Sbc         & 13.9 & 160 & 81 & 1 \\
UGC~12378   & NGC~7489 & 23 07 32.5 & 22 59 50 & Sd          & 14.0 & 164 & 78 & 1 \\
UGC~12486   & NGC~7591 & 23 18 16.2 & 06 35 09 & SBbc        & 13.9 & 167:& 61 & 1,2,8,18 \\
UGC~12539   & NGC~7631 & 23 21 26.7 & 08 13 05 & SA(r)b:     & 13.9 &  77 & 45 & 18 \\
UGC~12598   & NGC~7664 & 23 26 39.9 & 25 04 50 & Sc:         & 13.4 &  88 & 42 & 1,10,19 \\
UGC~12618   & NGC~7679 & 23 28 46.8 & 03 30 41 & SB0         & 13.5 & --  & 64 & 20 \\
UGC~12666   &          & 23 33 41.0 & 32 23 02 & Scd:        & 14.4 & 124 & 62 & 1 \\
UGC~12667   &          & 23 33 49.5 & 30 03 37 & Scd:        & 13.5 & 127 & 46 & 1 \\
UGC~12780   & NGC~7753 & 23 47 04.7 & 29 29 01 & SAB(rs)bc   & 12.8 &  71 & 64 & 21 \\
\noalign{\smallskip}
\hline
            &          &            &          &             &      &     &    & \\
\end{tabular}
\end{flushleft}

{\small Notes to Table~1.\\
Cols. 5,6: from the Third Refernce Catalogue of Bright Galaxies (RC3), de Vaucouleurs et al.
1991. Col. 7: from Moriondo et al. 1999.
Col. 9: references for the rotation curves. 1: Courteau 1992; 2: Mathewson et al. 1992;
3: Szomoru et al. 1994; 4: Afanas'ev 1988a; 5: Van Moorsel 1983; 6: Oosterloo \& Shostak 1993;
7: Vogt 1995; 8: Amram et al. 1994; 9: Broeils \& Van Woerden 1994; 10: Rubin et al. 1980;
11: M\'arquez \& Moles 1996; 12: Amram et al. 1992; 13: Blackman 1977; 14: Rubin et al. 1985;
15: Rubin et al. 1982; 16: Roelfsema \& Allen 1985; 17: Afanas'ev 1988b; 18: Rubin et al. 1988;
19: Rhee \& Van Albada 1996; 20: Jore 1997; 21: Marcelin et al. 1987.
}

% \resizebox{\hsize}{!}{\includegraphics*[10mm,60mm][200mm,270mm]{h1289t1.ps}}
\end{table*}

\section{The fits to the brightness distributions \protect\label{phot}}

\begin{figure*}
\vspace{1.5cm}
% \resizebox{\hsize}{!}{\includegraphics*[10mm,35mm][200mm,265mm]{h1289f1a.ps}}
\caption{Top panels: radial surface brightness profiles of the sample
galaxies, in a magnitude scale. For each plot, the triangles represent the
profile obtained by averaging the intensity over elliptical profiles (see
Moriondo et al. \cite{noi_1}); the dotted line is a cut along the major
axis of the galaxy; also shown are the best fit to the data (solid line) and 
the contributions from bulge (dot--dash line) and disk (dashed line). 
Bottom panels: the observed RC's (triangles) with the best fit to the data
(solid line). Again, the dot--dash and dashed lines represent the
contributions from bulge and disk respectively. The contribution from
the dark halo is shown as a dotted line.}
\label{fig:fit}\protect
\end{figure*}
 
\setcounter{figure}{0}
\begin{figure*}
\vspace{1.5cm}
% \resizebox{\hsize}{!}{\includegraphics*[10mm,35mm][200mm,265mm]{h1289f1b.ps}}
\caption{Continued}
\end{figure*}

The photometric data reduction and analysis are described in detail in 
Moriondo et al. (\cite{paper_1}, \cite{noi_1}). Briefly, 
each galaxy image is fitted with a model consisting of an exponential disk
and a bulge, whose shape is described by a generalized exponential (S\`ersic
\cite{sersic}). Both brightness distributions are assumed to be elliptical 
in the plane of the sky, and with a common centre and major axis. The 
parameters of the fit are the two scalelengths, surface brightnesses and 
apparent ellipticities (one for each component).
The results of these decompositions
were used in Moriondo et al. (\cite{noi_2}) to evaluate the effect
of internal extinction on the $H$-band structural parameters and derive 
the corrections to the face-on aspect. The corrected
values for scalelengths, surface brightnesses and total luminosities
will be used in the following analysis.
In most cases an exponential bulge yields a satisfactory fit to the data.
For two galaxies in the subsample considered here (namely UGC~26 and UGC~820), 
a value $n=2$ of the ``shape'' parameter in the exponent of the bulge 
brightness distribution produces better results. 
In two cases (UGC~673 and UGC~975) the disk alone is sufficient to fit 
the galaxy, i.e. no trace of a bulge is found in the brightness
distribution.
Table \ref{tab:phot} contains the photometric parameters of the sample
galaxies, corrected to the face--on aspect and for the redshift;
the $H-$band total luminosity, in column 8, will be introduced in Sect. 5.
The resulting decompositions are plotted in the top
panels of Fig.~\ref{fig:fit}, together with the surface brightness profiles 
averaged along elliptical contours (from Moriondo et al. 1999), and the 
brightness profiles along the major axis. In the case of UGC 673 and UGC
12666, the discrepancy between these two profiles is due to the 
ellipse--fitting routine, which in both cases has underestimated the galaxy 
apparent ellipticity in the regions of lower signal--to--noise ratio. 
Our fits, however, are in good agreement with the respective brightness 
distributions, so that the estimated parameters 
are reliable for these galaxies as well.

\begin{table*}
\caption{The photometric parameters}
\begin{flushleft}
\protect\label{tab:phot}
\begin{tabular}{cccccccc}
\noalign{\smallskip}
\hline
\noalign{\smallskip}
Name & $\mu_e$     & $R_e$ (kpc) & $\epsilon_b$ & $\mu(0)$    & $R_d$ (kpc) & $i$ (deg) & ${\cal M}_H$ \\
%%      & mag         & kpc   &              & mag         & kpc   & deg & mag   \\
%%      & arcs$^{-2}$ &       &              & arcs$^{-2}$ &       &     &       \\
 (1) & (2)         & (3)   & (4)          & (5)         & (6)   & (7) & (8)   \\
\noalign{\smallskip}
\hline
\noalign{\smallskip}
 UGC~00014 & 16.45 (0.10) &  0.66 (0.04) &  0.27 (0.01) & 17.76 (0.08) &  4.08 (0.18) & 44.0 ( 1.9) & -24.03 (0.11) \\
 UGC~00026 & 17.93 (0.11) &  0.94 (0.05) &  0.16 (0.02) & 19.51 (0.11) &  5.69 (0.34) & 55.9 ( 1.2) & -23.14 (0.13) \\
 UGC~00089 & 14.99 (0.10) &  0.75 (0.03) &  0.31 (0.02) & 16.72 (0.10) &  2.79 (0.12) & 56.0 ( 0.5) & -24.63 (0.08) \\
 UGC~00673 &  ...         &   ...        &   ...        & 18.10 (0.64) &  1.92 (0.67) & 58.7 ( 4.3) & -22.27 (0.94) \\
 UGC~00725 & 18.55 (0.33) &  0.35 (0.09) &  0.01 (0.23) & 18.20 (0.24) &  3.40 (0.30) & 76.9 ( 0.3) & -23.00 (0.17) \\
 UGC~00732 & 17.91 (0.23) &  0.40 (0.06) &  0.41 (0.01) & 17.84 (0.10) &  2.85 (0.12) & 55.2 ( 0.6) & -23.02 (0.08) \\
 UGC~00820 & 16.00 (0.26) &  0.81 (0.05) &  0.22 (0.01) & 16.82 (0.25) &  2.85 (0.18) & 67.0 ( 0.5) & -24.50 (0.23) \\
 UGC~00927 & 16.92 (0.13) &  0.29 (0.03) &  0.12 (0.08) & 18.03 (0.09) &  3.83 (0.15) & 54.8 ( 0.5) & -23.44 (0.07) \\
 UGC~00940 & 19.14 (0.21) &  0.68 (0.12) &  0.15 (0.17) & 18.86 (0.10) &  4.16 (0.20) & 49.2 ( 1.7) & -22.86 (0.11) \\
 UGC~00975 &  ...         &   ...        &   ...        & 17.84 (0.09) &  1.73 (0.06) & 52.4 ( 0.7) & -21.96 (0.07) \\
 UGC~01013 & 15.84 (0.14) &  1.14 (0.06) &  0.43 (0.01) & 17.74 (0.13) &  4.74 (0.25) & 61.4 ( 0.5) & -24.69 (0.10) \\
 UGC~01033 & 17.56 (0.38) &  1.15 (0.17) &  0.75 (0.01) & 18.53 (0.36) &  2.87 (0.38) & 81.3 ( 0.4) & -22.82 (0.33) \\
 UGC~01094 & 17.03 (0.23) &  0.66 (0.07) &  0.23 (0.04) & 17.50 (0.21) &  3.40 (0.28) & 74.3 ( 0.3) & -24.21 (0.15) \\
 UGC~01238 & 16.53 (0.22) &  0.85 (0.02) &  0.14 (0.01) & 18.08 (0.22) &  3.02 (0.11) & 42.3 ( 1.4) & -23.38 (0.23) \\
 UGC~01302 & 15.19 (0.16) &  0.32 (0.04) &  0.29 (0.07) & 17.48 (0.14) &  2.41 (0.15) & 63.4 ( 0.9) & -23.39 (0.12) \\
 UGC~01350 & 17.35 (0.23) &  0.94 (0.03) &  0.35 (0.01) & 18.34 (0.23) &  6.15 (0.22) & 51.1 ( 0.6) & -24.25 (0.22) \\
 UGC~01437 & 15.92 (0.09) &  0.54 (0.02) &  0.28 (0.01) & 16.39 (0.07) &  2.64 (0.07) & 45.1 ( 0.7) & -24.37 (0.06) \\
 UGC~01633 & 16.26 (0.16) &  0.47 (0.03) &  0.42 (0.05) & 16.63 (0.13) &  2.76 (0.14) & 61.6 ( 0.3) & -24.22 (0.10) \\
 UGC~01835 & 17.19 (0.12) &  0.75 (0.06) &  0.06 (0.01) & 17.80 (0.09) &  4.32 (0.23) & 49.8 ( 1.6) & -23.95 (0.12) \\
 UGC~01935 & 14.97 (0.21) &  0.50 (0.04) &  0.43 (0.02) & 17.18 (0.19) &  4.33 (0.33) & 72.3 ( 0.3) & -24.88 (0.13) \\
 UGC~01937 & 16.54 (0.18) &  0.58 (0.05) &  0.32 (0.06) & 16.97 (0.17) &  3.13 (0.12) & 52.3 ( 0.8) & -24.20 (0.16) \\
 UGC~02134 & 17.51 (0.17) &  0.95 (0.08) &  0.15 (0.05) & 17.81 (0.16) &  4.15 (0.31) & 66.4 ( 0.9) & -23.89 (0.15) \\
 UGC~02142 & 15.57 (0.25) &  1.01 (0.05) &  0.42 (0.01) & 17.79 (0.24) &  4.64 (0.25) & 61.6 ( 0.5) & -24.58 (0.23) \\
 UGC~02185 & 17.17 (0.29) &  0.22 (0.04) &  0.16 (0.19) & 17.81 (0.21) &  3.64 (0.30) & 74.8 ( 0.3) & -23.56 (0.14) \\
 UGC~02223 & 19.67 (0.77) &  0.40 (0.34) &  0.00 (0.86) & 18.26 (0.30) &  2.76 (0.29) & 74.3 ( 0.9) & -22.51 (0.29) \\
 UGC~02241 & 17.75 (0.23) &  1.93 (0.15) &  0.20 (0.03) & 18.24 (0.36) &  4.20 (0.66) & 50.7 ( 2.8) & -23.93 (0.47) \\
 UGC~02617 & 17.69 (0.20) &  0.52 (0.09) &  0.38 (0.14) & 18.20 (0.13) &  3.16 (0.20) & 60.9 ( 1.1) & -22.88 (0.13) \\
 UGC~02655 & 18.65 (0.15) &  1.93 (0.27) &  0.15 (0.07) & 18.97 (0.26) &  6.26 (1.07) & 62.6 ( 2.7) & -23.64 (0.40) \\
 UGC~02885 & 15.80 (0.17) &  0.69 (0.05) &  0.00 (0.04) & 17.96 (0.15) &  9.42 (0.59) & 67.0 ( 0.4) & -25.58 (0.10) \\
 UGC~11973 & 16.90 (0.29) &  0.35 (0.05) &  0.27 (0.16) & 16.84 (0.20) &  3.61 (0.27) & 72.4 ( 0.3) & -24.57 (0.14) \\
 UGC~12173 & 17.36 (0.26) &  0.38 (0.09) &  0.13 (0.25) & 17.80 (0.08) &  4.59 (0.19) & 49.4 ( 1.4) & -24.02 (0.08) \\
 UGC~12230 & 17.34 (0.17) &  0.49 (0.09) &  0.00 (0.14) & 17.00 (0.11) &  3.10 (0.15) & 57.9 ( 0.7) & -24.01 (0.09) \\
 UGC~12378 & 17.14 (0.17) &  0.42 (0.07) &  0.19 (0.18) & 17.51 (0.10) &  3.40 (0.15) & 54.9 ( 1.0) & -23.77 (0.08) \\
 UGC~12486 & 15.74 (0.11) &  0.66 (0.03) &  0.00 (0.02) & 17.92 (0.11) &  4.04 (0.22) & 56.8 ( 1.1) & -24.12 (0.11) \\
 UGC~12539 & 16.71 (0.17) &  0.32 (0.02) &  0.23 (0.04) & 17.10 (0.16) &  2.12 (0.12) & 64.7 ( 0.3) & -23.17 (0.12) \\
 UGC~12598 & 16.24 (0.15) &  0.23 (0.03) &  0.16 (0.10) & 16.22 (0.11) &  1.43 (0.06) & 58.6 ( 0.4) & -23.18 (0.08) \\
 UGC~12618 & 15.49 (0.97) &  0.65 (0.03) &  0.03 (0.05) & 16.81 (0.98) &  2.01 (0.08) & 22.6 ( 3.8) & -23.83 (0.98) \\
 UGC~12666 & 19.32 (0.24) &  1.46 (0.25) &  0.33 (0.08) & 19.60 (0.27) &  6.97 (1.70) & 74.3 ( 2.4) & -23.14 (0.51) \\
 UGC~12667 & 17.73 (0.29) &  0.28 (0.06) &  0.38 (0.01) & 17.94 (0.09) &  2.84 (0.13) & 53.1 ( 1.2) & -22.82 (0.08) \\
 UGC~12780 & 16.14 (0.05) &  0.61 (0.03) &  0.06 (0.05) & 17.23 (0.05) &  4.75 (0.11) & 39.8 ( 0.7) & -24.81 (0.04) \\
\noalign{\smallskip} 
\hline
        &              &              &              &              &              &             &         \\
\end{tabular} 
\end{flushleft}

{\small Notes to Table~2.\\
Col 2.: bulge effective surface brigtness, in mag arcsec$^{-2}$. Col. 3:
bulge effective radius. Col. 4: bulge ellipticity. Col. 5: disk central
surface brightness in mag arcsec$^{-2}$. Col 6: disk scalelength. Col. 7:
disk inclination. Col. 8: Total absolute magnitude.
}

% \resizebox{\hsize}{!}{\includegraphics*[10mm,70mm][200mm,260mm]{h1289t2.ps}}
\end{table*}

\section{The fits to the rotation curves \protect\label{kin}}

To estimate the contribution of bulge and disk to a given RC we 
use the information from the photometric data to predict the shape
of the RC's of the two components, and derive their $M/L$ ratios 
from a best fit to the observed velocity profile. The accuracy of this 
technique, which has been 
used for a long time by many authors (van Albada et al. \cite{vanalb},
Kent \cite{kent1}, Martimbeau et al. \cite{mart}, Moriondo et al.
\cite{paper_2}), is actually impaired by the 
scarce knowledge on the contribution of the dark component to 
the overall RC.  This becomes certainly important beyond the disk rotation
peak, at around two disk scalelengths; however, it might also be 
significant at smaller
radii, especially for low luminosity and low surface brightness galaxies --
as suggested by Persic et al. (\cite{persic}) and de Blok \& McGaugh 
(\cite{dbmg}) -- but also
in the case of bright spirals (see Bosma, \cite{bosma}, for a recent
review on this topic). Courteau and Rix (\cite{cou_rix}) and Bottema 
(\cite{botte}, \cite{botte2}), for
instance, estimate the disk contribution to be about 60\% of the overall 
rotation at 2.2 scalelengths, whereas different authors (e.g. Verheijen
\& Tully \cite{verhe}, Dubinski et al. \cite{dubin}, Gerhard
\cite{gerhard}, Bosma \cite{bosma}) support a ``maximum disk'' scenario in 
which such contribution is about 85\%, at least for bright galaxies. 
The shape of the dark matter distribution is rather uncertain as well: 
even the reliability of the distributions derived from numerical
simulations of structure formation (e.g. Navarro et al.
\cite{navarro}) is weakened by their apparent discrepancy with the observed
RC's of low surface brightness galaxies.
Since in this paper we are mainly interested in the properties of bulges
and disks, the choice of the dark halo distribution is not likely to be
important, at least
as long as the visible matter dominates the mass distribution in the inner
galaxy regions.

To have an idea of how the fit to the RC is influenced by the inclusion
in the model galaxy of a dark component, we perform two different fits
for each RC: one using only bulge and disk, and one including also the 
contribution from a dark halo. 
The expressions for the rotational velocity of an exponential disk and
a generic ellipsoidal bulge are reported in Moriondo et al. (\cite{paper_2}),
as well as two possible dark halo distributions, namely a constant 
density sphere and a pseudo--isothermal one (Kent, \cite{kent1}). 
The first halo distribution yields a linearly rising RC, and is 
an approximation of the latter as $R$ tends to zero; the rotation velocity 
associated to the pseudo--isothermal sphere, on the other hand, tends to a 
constant value as $R$ goes to infinity. 
This last distribution has been considered only for the 4 galaxies whose RC is 
well sampled in the outer, flat part (if this is not the case, and the 
asymptotic rotation velocity is not well defined, usually the two types
of dark matter halos yield the same result, in the sense that
only the linear part of the isothermal sphere is used by the minimization
routine that fits the data).

In general we find that if a dark component is included in the fit, it 
never turns out to be dominant in the inner part of the RC, within two
disk scalelengths; this is also true when the dark matter distribution 
is well constrained by a very extended RC. 
In other words, the solutions we obtain are usually not very different 
from the ``maximum disk'' ones, implying that,
% i.e. the ones in which the visible component 
% is {\it forced} to fit the RC in its rising part. Thus, 
in most cases, the visible part of the galaxy provides a good match 
to the observed RC; we will therefore assume that the ``maximum disk'' 
hypothesis is basically correct for our galaxies. 
An alternative scenario, however, will also be considered in 
Sect.~\ref{sect:ml}.
In a few cases the available RC is not extended enough to constrain 
even a constant density halo, and in these cases we use the results 
from the ``bulge + disk'' fits.
We consider these values reliable, since in general the inclusion of 
the dark halo in the fits does not produce major changes in the estimated 
mass and $M/L$ of the visible components. 
% and slso because 
% the scatter in the $M/L$ values is not correlated with the relative
% contribution of the disk to the observed RC. In particular, for some
% galaxies we obtain a high $M/L$ for a disk which contributes 70\% of the 
% peak rotation, whereas for some others we find a low $M/L$ for a full 
% maximum disk solution.

We also note that, in a few cases, at 2.2 $R_d$'s the contribution to the
overall rotation from the bulge is still significant, so that the 
disk contribution is well below the measured circular velocity.
In these cases neglecting the presence of the bulge would 
certainly lead to overestimate the disk mass and $M/L$.

Table \ref{tab:dyn} contains the parameters derived from the RC fits.
Also included in the table are the values 
of the velocity width $W$ (corrected for instrumental smoothing,
redshift, turbulence, and inclination), and of the velocity at 2.2 $R_d$, 
$V_2$ (corrected for redshift and inclination); both quantities will be 
introduced in the 
next section. In Fig.~\ref{fig:fit} (bottom panels) we show the best fits 
to the RC's for all the galaxies in our sample.

\begin{table*}
\caption{The dynamical parameters}
\begin{flushleft}
\protect\label{tab:dyn}
\begin{tabular}{ccccccc}
\noalign{\smallskip}
\hline
\noalign{\smallskip}
Name & $M_b$            & $(M/L)_b$ & $M_d$            & $(M/L)_d$ & $W$ (km s$^{-1}$) & $v_2$ (km s$^{-1}$) \\
%%      & 10$^9 M_{\odot}$ &           & 10$^9 M_{\odot}$ &           & km s$^{-1}$ & km s$^{-1}$ \\
 (1) & (2)              & (3)       & (4)              & (5)       & (6)         & (7)         \\
\noalign{\smallskip}
\hline
\noalign{\smallskip}
 UGC~00014 &  12.2 ( 3.0) &  0.95 (0.25) &  97.5 ( 9.4) &  1.25 (0.16) & 456.7 (18.0)& 215.0 (18.2)  \\
 UGC~00026 &  ...         &  ...         &  22.5 ( 7.0) &  0.74 (0.25) & 271.6 ( 5.8)&  90.0 (27.6)  \\
 UGC~00089 &  21.7 ( 4.4) &  0.34 (0.08) &  29.2 ( 5.2) &  0.31 (0.06) & 445.2 ( 5.5)& 180.0 (31.2)  \\
 UGC~00673 &  ...         &  ...         &   8.9 ( 3.7) &  0.70 (0.53) & 299.2 (15.1)& 120.0 (26.0)  \\
 UGC~00725 &  ...         &  ...         &  24.5 ( 7.2) &  0.68 (0.26) & 293.1 ( 8.2)& 160.0 (44.9)  \\
 UGC~00732 &  ...         &  ...         &  17.4 ( 2.8) &  0.49 (0.09) & 289.7 (24.4)& 120.0 (18.7)  \\
 UGC~00820 &  ...         &  ...         &  55.1 ( 5.3) &  0.61 (0.16) & 462.7 ( 4.9)& 240.0 (17.4)  \\
 UGC~00927 &  ...         &  ...         &  34.7 ( 2.9) &  0.64 (0.08) & 314.3 (35.0)& 150.0 (11.2)  \\
 UGC~00940 &  ...         &  ...         &  65.7 ( 4.8) &  2.22 (0.27) & 328.8 (15.7)& 160.0 ( 8.9)  \\
 UGC~00975 &  ...         &  ...         &  13.1 ( 1.3) &  1.01 (0.13) & 275.0 ( 6.8)& 128.0 (12.0)  \\
 UGC~01013 & 181.1 (20.1) &  2.64 (0.46) &  81.1 (14.0) &  0.75 (0.16) & 550.9 ( 5.5)& 300.0 (49.5)  \\
 UGC~01033 &  10.2 ( 8.3) &  0.72 (0.64) &  12.7 ( 2.4) &  0.67 (0.28) & 323.1 ( 3.3)& 160.0 (22.0)  \\
 UGC~01094 &   3.9 ( 1.6) &  0.51 (0.25) &  49.2 ( 4.3) &  0.71 (0.16) & 435.2 ( 4.5)& 212.0 ( 5.6)  \\
 UGC~01238 &  23.8 ( 1.5) &  1.18 (0.27) &  28.7 ( 3.5) &  0.91 (0.23) & 397.9 (12.6)& 180.0 (21.1)  \\
 UGC~01302 &   9.0 ( 2.4) &  0.92 (0.29) &  23.1 ( 3.3) &  0.65 (0.13) & 353.9 ( 5.3)& 190.0 (24.8)  \\
 UGC~01350 &  ...         &  ...         & 169.2 (33.2) &  1.63 (0.49) & 731.3 (27.2)& 240.0 (46.2)  \\
 UGC~01437 &   6.6 ( 1.8) &  0.46 (0.13) &  49.6 ( 5.2) &  0.43 (0.05) & 420.1 ( 7.5)& 205.0 (20.6)  \\
 UGC~01633 &   4.5 ( 1.7) &  0.58 (0.24) &  56.9 ( 5.2) &  0.56 (0.09) & 479.4 ( 6.4)& 240.0 (18.3)  \\
 UGC~01835 &  10.2 ( 2.8) &  1.20 (0.36) &  65.4 (11.4) &  0.77 (0.15) & 423.2 (11.7)& 207.0 (34.2)  \\
 UGC~01935 &   1.4 ( 2.0) &  0.05 (0.07) &  46.8 ( 7.4) &  0.31 (0.08) & 447.0 ( 3.5)& 200.0 (27.8)  \\
 UGC~01937 &  ...         &  ...         &  56.5 ( 9.7) &  0.60 (0.14) & 437.8 ( 6.6)& 200.0 (33.5)  \\
 UGC~02134 &  15.5 ( 3.1) &  1.56 (0.41) &  34.8 ( 4.0) &  0.45 (0.09) & 335.6 ( 4.4)& 160.0 (13.9)  \\
 UGC~02142 &  65.1 (11.8) &  0.95 (0.29) & 109.2 ( 9.2) &  1.12 (0.29) & 520.1 ( 5.2)& 250.0 (16.3)  \\
 UGC~02185 &  ...         &  ...         &  37.6 ( 4.1) &  0.63 (0.15) & 437.1 (14.9)& 200.0 (14.2)  \\
 UGC~02223 &  ...         &  ...         &   7.2 ( 1.3) &  0.32 (0.11) & 244.9 (10.2)& 125.0 (19.5)  \\
 UGC~02241 &  82.3 (12.5) &  2.46 (0.68) &  44.6 (18.0) &  0.84 (0.45) & 461.9 (20.4)& 250.0 (93.0)  \\
 UGC~02617 &  ...         &  ...         &  22.4 ( 2.4) &  0.72 (0.12) & 535.4 (19.7)& 138.0 (12.0)  \\
 UGC~02655 &   9.1 ( 2.5) &  0.62 (0.20) &  39.1 ( 7.2) &  0.65 (0.20) & 307.0 (20.0)& 140.0 (10.2)  \\
 UGC~02885 &  53.8 ( 5.9) &  2.12 (0.43) & 185.7 (21.7) &  0.54 (0.10) &  ...        & 280.0 (27.6)  \\
 UGC~11973 &  ...         &  ...         & 104.1 (18.7) &  0.73 (0.20) & 503.2 ( 5.9)& 360.0 (59.1)  \\
 UGC~12173 &   5.2 ( 1.7) &  2.80 (1.16) & 104.2 ( 6.4) &  1.10 (0.11) & 465.3 (12.6)& 210.0 ( 9.7)  \\
 UGC~12230 &  ...         &  ...         &  83.1 ( 5.8) &  0.92 (0.12) & 517.0 ( 8.5)& 260.0 (12.9)  \\
 UGC~12378 &   1.3 ( 1.7) &  0.47 (0.60) &  52.3 ( 6.1) &  0.77 (0.12) & 407.4 ( 6.6)& 200.0 (21.5)  \\
 UGC~12486 &  33.1 ( 6.2) &  1.32 (0.29) &  61.9 ( 9.6) &  0.93 (0.18) & 432.6 ( 7.1)& 210.0 (30.5)  \\
 UGC~12539 &   2.2 ( 1.5) &  0.92 (0.62) &  21.5 ( 3.3) &  0.56 (0.12) & 375.2 ( 5.3)& 160.0 (22.5)  \\
 UGC~12598 &  ...         &  ...         &  18.8 ( 1.5) &  0.48 (0.06) & 360.2 ( 4.0)& 170.0 (11.0)  \\
 UGC~12618 &  14.1 ( 4.7) &  0.47 (0.48) &  36.2 (14.0) &  0.80 (0.84) & 630.9 (105.4)& 215.0 (82.6)  \\
 UGC~12666 &   3.8 ( 1.6) &  0.84 (0.41) &  47.0 (11.7) &  1.12 (0.41) & 271.9 ( 5.5)& 150.0 ( 7.1)  \\
 UGC~12667 &  ...         &  ...         &  21.5 ( 2.4) &  0.67 (0.09) & 253.7 ( 6.0)& 140.0 (14.2)  \\
 UGC~12780 &   1.3 ( 6.2) &  0.09 (0.42) & 161.0 (12.2) &  0.94 (0.08) & 515.8 (10.0)& 260.0 (18.9)  \\
\noalign{\smallskip} 
\hline
           &              &              &              &              &              &              \\
\end{tabular} 
\end{flushleft}

{\small Notes to Table~3.\\
Cols. 2,4: masses in units of 10$^9$ M$_{\odot}$. Cols. 3,5: M/L's in solar units.
}

% \resizebox{\hsize}{!}{\includegraphics*[10mm,70mm][200mm,260mm]{h1289t3.ps}}
\end{table*}

In 17 cases out of 40, the observed RC has either no data-points in the
innermost region (3 cases), or shows no evidence of contribution by 
the bulge. For these galaxies the bulge mass and $M/L$ are either 
unconstrained or likely to be underestimated. 
Most of the 14 galaxies, for which the disk contribution alone is sufficient 
to match the inner RC, are
characterized by the smallest bulge--to--disk ratios ($B/D$) of the sample 
($B/D < 0.05$), or by the faintest bulge luminosities (${\cal M}_b > -21$). 
It is not surprising, therefore, that in these cases the bulge contribution
to the overall RC cannot be adequately resolved.
A direct comparison between the $M/L$ of disks and bulges seems to 
confirm these statements. In Fig.~\ref{fig:mllbd} mass 
versus luminosity is plotted for the disks, and for the bulges with 
$M_b > 10^9 M_{\odot}$. These are the 23 cases for which we obtain a 
reliable fit to the RC for both components, and the two sets of objects seem 
to form a smooth sequence, roughly delimited by $M/L = 0.25$ and 
$M/L = 2.5$, in solar units.
% (the two apparently discrepant data points have large errorbars and are not 
% inconsistent with the rest of the sequence). 
The remaining ``low mass'' 
bulges are placed well below the sequence, out of the plot window, suggesting
that their mass is largely underestimated. 
In the following analyisis, therefore, we will consider only the 
bulges plotted in Fig.~\ref{fig:mllbd}.

\begin{figure}
\vspace{1.5cm}
% \resizebox{\hsize}{!}{\includegraphics*[5mm,55mm][190mm,235mm]{h1289f2.ps}}
\caption[]{Mass versus luminosity plotted for bulges and disks of our
sample galaxies. Open circles represent the bulges, whereas filled triangles
represent the disks.  Only bulges with mass $> 10^9 M_{\odot}$ are plotted.
The slope of the two dashed line is of constant $M/L$. The two $M/L$ values 
plotted, 2.5 and 0.25, are in solar units ($H$ band).}
\label{fig:mllbd}\protect
\end{figure}

\section{The Tully Fisher relation for the sample \protect\label{sec:tf}}

To derive an $H$-band TF relation for our galaxies the 21~cm line velocity 
widths from the RG and MH database (Table 3) have been corrected for instrumental
smoothing, redshift, turbulence, and inclination of the galaxy to the 
line of sight, according to the prescriptions in Giovanelli et al. 
(\cite{giova97}, \cite{giovanew}). The absolute galaxy magnitudes in the $H$ 
band (Table 2) are derived integrating the surface brightness profiles 
extrapolated up to 8 disk scalelengths; a small correction to face-on aspect 
is applied, using the results obtained for the disks alone in Moriondo et 
al. (\cite{noi_2}). 
Figure \ref{fig:tf} shows the plot we obtain, with the best fit to the 
data after the exclusion of the three discrepant points to the right. 
In the case of UGC~12618, our value for the inclination is probably 
underestimated, leading to an exceedingly large correction for $W$; on 
the other hand, due to its large errorbars, the contribution of this galaxy 
to the fit is not very significant anyway. In the case of UGC~1350
and UGC~2617, the observed $W$ is likely to be overestimated, maybe for a 
misidentification of the galaxy; we note that for these two objects 
an estimate of the rotation derived from the RC is, respectively, about 75\% 
and 50\% smaller than $W$. 

The errors on both axes are of comparable magnitude, therefore the weight for 
each data point is the RMS of the two contributions.
Since the relation we fit is $y = ax + b$, the i-th residual is weighted by
\begin{equation}
\protect\label{eq:wei}
 w(i) = \left [\sqrt{\epsilon_{y(i)}^2 + a^2 \epsilon_{x(i)}^2}\right ]^{-1} \;\;,
\end{equation}
where the $\epsilon$'s are the estimated errors.
The slope and zero offset, with respect to the average $\log W$, are
respectively $-7.7 \pm 1.0$ and $-23.67 \pm 0.08$.

\begin{figure}
\vspace{1.5cm}
% \resizebox{\hsize}{!}{\includegraphics*[5mm,55mm][190mm,235mm]{h1289f3.ps}}
\caption[]{The $H$-band Tully-Fisher relation for our sample, with the 
best fit to the data after the exclusion of the three discrepant points 
on the right side. Different symbols correspond to different morphological
types: filled circles for Sa-Sab, filled squares for Sb-Sbc, open squares
for Sc-Scd, crosses for Sd and dm.}
\label{fig:tf}\protect
\end{figure}

\section{Fundamental Planes}

We now fit separately to bulges and disks a relation 
\begin{equation}
\protect\label{eq:fp}
\log R = a \log V + b \log I + c
\end{equation} 
involving a scalelength, a velocity, and a surface brightness, analogous
to the FP of elliptical galaxies.
The designated parameters for the bulges are the effective 
radius $R_e$, the surface brightness at $R_e$ ($I_e$),
and a velocity defined as 
\begin{equation}
\protect\label{eq:v_b}
V_b = \sqrt{\frac{M_b}{R_e}}
\end{equation}
where $M_b$ is the bulge mass, in units of 10$^9 M_{\odot}$.
Since for all the bulges selected (the ones with $M_b > 1$) the 
brightness distribution is fitted by a pure exponential -- i.e. 
they all have the same shape -- this velocity scales as the rotation velocity,
measured at 2.2 bulge scalelengths $R_b$. Also, $R_e$ and $I_e$ are related 
by constant factors respectively to $R_b$
and the central surface brightness $I(0)$ (the two parameters usually adopted 
for the disks): $R_e = 1.67\; R_b$, and $I_e = 0.19\; I(0)$.
For the disks we choose the exponential scalelength $R_d$, the central surface 
brightness $I(0)$, and a velocity $V_d$ defined via a relation analogous to Eq.
(\ref{eq:v_b}), with the disk mass and the disk scalelength. Again, this 
velocity differs from the value at 2.2 $R_d$ by the same factor for
all the exponential distributions.
Following CY95, we also consider a relation involving $R_d$, $I(0)$, and the 
{\it total galaxy rotation velocity} at 2.2 $R_d$, derived directly from the
RC, which we will indicate as $V_2$. 

It turns out that the three parameters to be fitted have comparable 
uncertainties, and each point needs to be weighted by a combination of them.
In the case of the structural parameters ($R$, $I$), besides the 
formal error from the fit to the brightness distributions, a major 
contribution
is added from the errors in the corrections to face-on aspect, 
that is from the uncertainty in the amount of internal extinction. 
These are evaluated according to the results in Moriondo et al. (\cite{noi_2}).
The uncertainty on the velocity depends in principle on the 
errors associated both to the RC measurement and to the scale length. 
However we expect this latter contribution to be less important, 
especially for the disks, whose scalelength is always the best determined 
parameter in the surface brightness decompositions. Therefore we assume 
the uncertainty on the velocities to be well represented by the formal error 
derived from the fitting routine. 
The errors on all other quantities are derived from these
values: for example, the error on the disk mass is obtained combining the 
uncertainties on the velocity and the scale length, since 
$M \sim V^2\, R$. 

The fit to the disk parameters yields
\begin{equation}
\protect\label{eq:fpd}
\begin{array}{lll}
\log R_d & =  & (1.31\, \pm\, 0.19) \, (\log V_d - <\log V_d >) + \\
         &    & (-0.62\, \pm\, 0.09) \, (\log I(0) - <\log I(0) >) + \\
         &    &  (1.29\, \pm\, 0.07) \;\;\; , \\
\end{array}
\end{equation}	
where ``$<\; >$'' designates the average over the sample.
If $V_2$ is chosen, instead of the value defined in terms of the disk mass and 
scalelength, we obtain
\begin{equation}
\protect\label{eq:fpd2}
\begin{array}{lll}
\log R_d & = & (1.47\, \pm\, 0.16) \, (\log V_2 - <\log V_2 >) + \\
         &   & (-0.61\, \pm\, 0.07) \, (\log I(0) - <\log I(0) >) + \\
         &   & (1.39\, \pm\, 0.06) \;\;\; . \\
\end{array}
\end{equation}	
We note that these coefficients are close to the values implied 
by the TF relation derived in Sect. \ref{sec:tf}:
in fact, from $L \sim W^{3.1}$, we derive $R \sim W^{1.5}I^{-0.5}$.
Therefore, the TF relation is a nearly edge-on 
projection of the FP. Also, these values are not inconsistent with 
the constraint implicit in the relation suggested by CY95 
(Eq. (\ref{eq:cy}) ), i.e. $a = 2b$.
% Also, it appears that the scaling of the 21~cm line velocity width is not 
% very 
% different from the scaling of the rotation velocity measured at 2.2 $R_d$
% (a conclusion also implicit in the results by Giovanelli et al. 
% \cite{giovanew}). 

The best fit to the bulge parameters yields:
\begin{equation}
\protect\label{eq:fpb}
\begin{array}{lll}
\log R_e & = & (0.97\, \pm\, 0.13 ) \, (\log V_b - <\log V_b >) + \\
         &   & (-0.61\, \pm\, 0.08) \, (\log I_e - <\log I_e >) + \\
         &   & (-0.46\, \pm\, 0.10) \;\;\; . \\
\end{array}
\end{equation}	
Again, a relation like Eq. (\ref{eq:cy}) is not ruled out, even 
if with a slope $a$ slighlty different from the disks' value.

The errors associated to the various coefficients are estimated by performing 
a large number of Monte Carlo simulations of the data sample and fitting 
each simulation, to derive a distribution of values for each coefficient. 
Figures \ref{fig:fpd} and \ref{fig:fpd2} show the relations we find for 
the disks, whereas Fig. \ref{fig:fpb} shows the one for the bulges.

\begin{figure}
\vspace{1.5cm}
% \resizebox{\hsize}{!}{\includegraphics*[5mm,55mm][190mm,235mm]{h1289f4.ps}}
\caption[]{The scaling relation of Eq. (\ref{eq:fpd}) for the disks, 
derived from the best fit to the RC. The data and the best fit 
are shown. 
The factorization of the $I(0)$ coefficient in the $x-$axis label 
is chosen to make it easily comparable with Eq. (\ref{eq:cy}). 
Symbols are as in Fig. \ref{fig:tf}.}
\label{fig:fpd}\protect
\end{figure}

\begin{figure}
\vspace{1.5cm}
% \resizebox{\hsize}{!}{\includegraphics*[5mm,55mm][190mm,235mm]{h1289f5.ps}}
\caption[]{The scaling relation of Eq. (\ref{eq:fpd2}) for the disks,
derived using the rotation velocities at 2.2 $R_d$. The data and the best
fit are shown. Symbols as in Fig. \ref{fig:tf}.}
\label{fig:fpd2}\protect
\end{figure}

\begin{figure}
\vspace{1.5cm}
% \resizebox{\hsize}{!}{\includegraphics*[5mm,55mm][190mm,235mm]{h1289f6.ps}}
\caption[]{The scaling relation of Eq. (\ref{eq:fpb}) for the bulges.
The data and the best fit are shown. Symbols as in Fig. \ref{fig:tf}.}
\label{fig:fpb}\protect
\end{figure}

\subsection{A comparison with published results}

The coefficients we derive for the disks are different
from the values found by CY95 ($a = 1.0$, $b = -0.5$), but closer to the
ones reported by Karachentsev (\cite{kara}), i.e. $a = 1.4$, $b = -0.7$; 
we note however that, in the case of CY95, the photometric data 
were in a very different passband ($B$).  
Both our coefficients and the other quoted values are not consistent 
with what would be expected on the basis of the virial theorem and a 
universal mass--to--light ratio ($a=2$ and $b=-1$).

The various sets of coefficients can also be compared to the ones which 
define the FP of elliptical galaxies, evaluated using 
the central velocity dispersion as the kinematic parameter, and 
the effective surface brightness.  
For example, Bender et al. (\cite{bbf}) report for the Virgo cluster
$a = 1.4$ and $b = -0.85$, in the $B$ band; more recently J\o rgensen et al. 
(\cite{jorg}) find $a = 1.24$ and $b = -0.82$ in the Gunn $r$ band, with a 
scatter of 
0.084 in $\log R_e$; Pahre et al. (\cite{pahre}) estimate the NIR
coefficients of the plane to be $a = 1.53$ and $b = -0.79$.
The various FP's are not very different, and actually the 
existence of a ``cosmic metaplane'' has already been claimed by Bender et al. 
(\cite{bendk}), and BBFN.
In particular they defined a set of three parameters (the $k$ parameters) 
particularly suited to
represent the FP of elliptical galaxies in the $B$ band, and found 
that basically all the self--gravitating systems show a similar behaviour
in the $k$-parameter space.
In the case of spiral galaxies, they considered their 
global properties, without attempting a decomposition into structural 
components, and for the bulges in their data sample they used the 
central velocity dispersion as the kinematical parameter.
This work improves the approach by characterizing separately bulges 
and disks from the photometric point of view; in addition, we are 
able to determine the bulge mass independently of its kinematical status 
(i.e., if it's more or less supported by rotation), and obtain an
independent estimate of the coefficients of the bulge FP. 

Figure \ref{fig:ksp} shows our data in the $H-$band $k$--space, with 
bulges denoted as open circles, and disks as triangles. 
We have defined the three coordinates as
$k_1 = \log(M)$, $k_2 = log(M/L \;\cdot\; I_e^3)$, and $k_3 = \log(M/L)$, 
where $M$ is the total mass in units of $M_{\odot}$, $M/L$ is the
stellar mass--to--ligth ratio in solar units, and $I_e$ is the effective 
surface brightness in $L_{\odot}\, pc^{-2}$. These definitions, besides 
being applied to a different passband, are slightly
different from the ones introduced by Bender et al. (\cite{bbf}).
We can calculate the transformations between the two sets of
coordinates from the relations reported in the Appendix A
to BBFN, and assuming that all our bulges and disks are adequately
described by exponentials. Using a typical $B-H = 3.5$ for both components, 
we find

\vspace{0.2cm}
\begin{equation}
\protect\label{eq:trans}
\left\{ \begin{array}{lll}
		\; k^B_1 & = & \frac{1}{\sqrt{2}}\; (k_1 - 5.97) \\
		\; k^B_2 & = & \frac{1}{\sqrt{6}}\; (k_2 + 0.58) \\
		\; k^B_3 & = & \frac{1}{\sqrt{3}}\; (k_3 + 1.37) \\
	  \end{array} \right.
\end{equation}
\vspace{0.2cm}
\noindent
where $k^B_1$, $k^B_2$, and $k^B_3$ are the $B-$band BBFN parameters.

The dotted line in the upper right corner of the $k_1$ vs. $k_2$ plot 
corresponds, in our set of coordinates, to the boundary of the so--called 
Zone of Exclusion (ZOE), defined in the $B$ band by $k^B_1 + k^B_2 > 8$:
it is consistent with our data, in the sense that most bulges and all 
the disks are placed to its left side. 
The BBFN database for the bulges is here extended to lower 
masses, with several data--points falling in the typical range of dwarf 
ellipticals
($M < 10^{10} M_{\odot}$); the two classes of objects appear however
separated, with the bulges shifted towards higher values of $k_2$, and 
therefore higher concentrations (note that, for a given mass,
$k_2$ is proportional to $\log (I_e^2/R_e^2)$, which is higher for compact,
bright objects).

For what concerns spiral galaxies, BBFN find that their
average distance from the ZOE increases steadily with morphological type,
from Sa's to Irregulars;
the data points of our disks, however, are all
placed in about the same region of the $k_1$--$k_2$ plane, quite distant
from the ZOE, and roughly coincident with the locus occupied by Scd's 
galaxies in the BBFN's plots.
Most likely, this difference arises from the separation of the two structural 
components, which has allowed us to plot the $k$ parameters of the {\it 
disks alone}: if the values for the whole galaxies 
are considered, one would expect the systems with higher
$B/D$ (namely, the early--type spirals) to lie closer to 
the ZOE, due to the contamination of the bulge. The BBFN sequence, therefore, 
would be mainly driven by the average $B/D$, which in turn is roughly
correlated with morphological type.
When considered separately, on the other hand, disks and bulges are
located in two distinct, contiguous regions of the $k-$space, as it 
is also evident from the $k_2$--$k_3$ projection, with the disks shifted 
towards higher masses and lower concentrations.
A different ZOE could in principle be defined for the disks, with about the 
same slope but shifted by about two decades towards lower $k_2$ values.

In the top panel of Fig. \ref{fig:ksp}, we have plotted the slope of the 
$B-$band FP for 
the Virgo cluster, scaled to the centroid of the bulges (dashed line) 
and of the disks (solid line).
It appears to be consistent with our data, as suggested by the similarity
of our coefficients in Eqs. (\ref{eq:fpd}) and (\ref{eq:fpd2}) to the ones 
reported by Bender et al. (\cite{bbf}). 
Also, using Eqs. (\ref{eq:trans}) and the definition in BBFN, we can 
estimate the 
quantity $\delta_{3:1}$ for bulges and disks, representing their average
vertical distance from the ellipticals' FP in the $k^B_1$--$k^B_3$ projection.
In the case of the bulges, we find $\delta_{3:1} = 0.03$ dex, in fair
agreement with the BBFN estimate (-0.03); for the disks
we find a shift of -0.19 dex, which is about the value found by BBFN for 
spirals from type Sa to Sc. 

%The distance between the two lines amounts to 0.02 in 
%$\log M/L$; since the scatter of the data points is larger by about 
%one order of magnitude, we conclude that no significant difference is 
%detected, for this sample, between the average $H-$band $M/L$ of bulges 
%and disks.
% We can attempt a rough comparison of this discrepancy 
% with the predictions of stellar population synthesis (SPS) models. 
% Using the results by Worthey (\cite{worthey}) and Bruzual \& Charlot 
% \cite{br_ch} for single--burst stellar populations, we find that a factor 
% of about 1.5 in $M/L$ can 
% be produced by a different {\it age} of the average stellar population of
% bulges and disks (in particular, a factor of about 1.5 is necessary, 
% as, for example, in the case of disks about 8 Gyr old, and bulges 12 Gyr old)
% but, according to Worthey's models, also a disk {\it metallicity} 
% $\sim$6 times higher than the bulge value would yield a similar 
% difference in $M/L$. An {\it IMF} more biased towards low-mass stars 
% for the bulges would also be effective, as the Bruzual \& Charlot models 
% demonstrates, or some combination of all these factors. 
% As a matter of fact, comparing $M/L$ values in a single band is not sufficient
% to distinguish between the different possibilities; some additional 
% information, however, might be provided by a detailed multiband analysis, 
% -- possibly integrated by spectroscopic data -- as the one recently attempted
% by Pahre et al. (\cite{pahre2}) for elliptical galaxies. 

\begin{figure}
\vspace{1.5cm}
% \resizebox{\hsize}{!}{\includegraphics*[20mm,70mm][175mm,215mm]{h1289f7.ps}}
\caption[]{Bulges (open circles) and disks (triangles) in the $k-$space
defined by Bender et. al (\cite{bbf}). In the top panel the slope of the
FP for elliptical galaxies is plotted for comparison. In the $k_1$ versus
$k_2$ panel the dotted line represents the Zone of Exclusion,
as defined by BBFN.}
\label{fig:ksp}\protect
\end{figure}

\section{The disk FP as a distance indicator}

In order to assess the goodness of Eqs. (\ref{eq:fpd}) and (\ref{eq:fpd2}) 
as tools to
measure galaxy distances, we compare the scatter of the data points, with 
respect to the best fit, to the scatter associated to the TF relation.
The dispersion is defined as 
\[ \sigma = \sqrt{\frac{\sum_i{(\Delta_i w_i)^2}}{\sum_i{w_i^2}}} \;\; ,\]
where $\Delta_i$ is the residual on the i-th data point and $w_i$ the 
associated weight, defined by Eq. (\ref{eq:wei}). 
We find a dispersion of 0.38 mag for the TF relation, implying an uncertainty 
of 19\% on the distance of a single galaxy. We note that the galaxy 
distances used in this context have been estimated purely from redshifts.
The peculiar velocity field thus adds scatter to the corresponding TF
relation in a measure which we estimate on the order of 0.15 to 0.20 mag,
at the typical distance of these objects. Had peculiar velocity--corrected
distances been used, the TF scatter would have been so that distance
estimates would have uncertainties of about 16\%, rather then 19\%.

In the case of the FP, we find a 
dispersion of 0.11 in $\log r$ if we use velocities derived from the best 
fits (Eq. (\ref{eq:fpd})), and of 0.09 if the velocities 
are the actual rotation velocities (Eq. (\ref{eq:fpd2}) ). 
The associated uncertainties in the 
distance are respectively about 29\% and 23\%.
Even if these two values are less than 1/2 of the dispersion quoted by 
Giovanelli for the relation proposed by CY95 and a sample of 153 spiral 
galaxies (0.25 in $\log r$, not based on a detailed disk--bulge
decomposition analysis as we use here), they are still larger than the 
uncertainty yielded by the TF relation for our sample.
Therefore, according to our data, the disk FP is not as accurate,
as a distance indicator, as the TF relation, even if it allows for one 
more free parameter in the relation. 
This result could be imputed to the scatter added to the relation by the 
uncertainties introduced in the data analysis process (in particular,
the decomposition of the surface brightness distribution and the fit to the 
RC). On the other hand, we find that a modified Tully-Fisher relation,
in which the velocity width is replaced by $V_2$, is characterized as well 
by a larger dispersion of 0.45 mag. This value is equivalent to the 
dispersion of 0.09 in $\log r$ associated to Eq. (\ref{eq:fpd2}). 
In a similar way, if the total luminosity is replaced 
by the luminosity within 2.2 $R_d$'s, we find a very large dispersion of 
0.64 mag. 
Since these alternative parameters are not strictly derived from the 
surface brightness decompositions or by the fit to the RC's, as it is the
case for the FP ones, we argue that using parameters associated with the 
inner part of the galaxy (rather than global quantities), is in itself a 
major source of scatter in the scaling relation considered.
Thus, the fine tuning between photometric and kinematic
parameters that yield the TF relation would be more effective because
of their global character.
About the much reduced dispersion with respect to the Giovanelli data,
we attribute this difference to the use of a more refined
data-analysis technique for our sample.

\section{The relation between mass and luminosity \protect\label{sect:ml}}

Eqs.~(\ref{eq:fpd})--(\ref{eq:fpb}) show that for both disks and bulges 
a systematic variation of $M/L$ exists,
if we assume that we have properly evaluated their
shape, photometric properties (including the effect of internal 
extinction), and contribution to the RC.
In particular we can rewrite Eqs. (\ref{eq:fpd}) and (\ref{eq:fpb}) as
\begin{equation}
\protect\label{eq:lmd}
L_d \sim \frac{M_d^{1.0 \pm 0.2}}{R_d^{0.7 \pm 0.4}} 
%L_d \sim \frac{M_d^{1.0}}{R_d^{0.7}} 
\end{equation}
and
\begin{equation}
\protect\label{eq:lmb}
L_b \sim \frac{M_b^{0.8 \pm 0.2}}{R_e^{0.5 \pm 0.3}} \;\; .
%L_b \sim \frac{M_b^{0.8}}{R_e^{0.5}} \;\; .
\end{equation}
Whereas for elliptical galaxies the $M-L$ plane shows an edge-on view of 
the FP, this is not the case for disks and bulges, and a residual dependence
on the scalelength is left; this dependence is such that, if a given mass 
settles into a larger size, the corresponding $M/L$ ratio is larger. 
Actually, this result is already implicit in the work by CY95: a good match
to equation  \ref{eq:cy} with $a \neq 2$ is possible only if $M/L$ is a
function of $R_d$ alone.
Previous results, both observational (Persic et al. 1996), and
theoretical (Navarro et al. \cite{navarro}, Navarro \cite{nava}), have 
suggested that the stellar
$M/L$ in the $B$ band increases with the galaxy luminosity, rather than
with its size. This conclusion is probably consistent with ours, since
the total galaxy luminosity, especially in the $B$ band, is mainly
determined by the disk scalelength. 
Since the average stellar $M/L$ ratios are determined by the
star--formation history of the galaxy, a connection must necessarily 
exist between the structural properties of the system, in
particular its size, and the characteristics of its average stellar
population. This is not surprising, since connections of this kind also 
exist for ellipitical galaxies: for example the one at the base of the 
well--known color--magnitude relation. 
On the other hand, the parameters which characterize stellar populations
(age, metallicity, star--formation rate, etc.), whose systematic variation
could be responsible of the observed trend, are far too many to be
constrained by our $H-$band data alone.
Some additional information on this issue could be provided by a detailed 
multiband 
analysis -- possibly integrated by spectroscopic data -- as the one recently
carried out by Pahre et al. (\cite{pahre2}) for elliptical galaxies.

%In this case, however, the size alone is involved: this implies that 
%two galaxies with the same luminosity but different $R_d$ are expected to 
%have different $M/L$'s. Moreover, a corresponding color--size relation has 
%not been observed (see for example
%Courteau and Rix \cite{cou_rix}), so that the systematic variation in $M/L$ 
%implied by Eqs. (\ref{eq:lmd}) and (\ref{eq:lmb}) will be roughly the same 
%at all wavelengths. 

We note however that a different scenario can be outlined, in particular
if we forget Eqs. (\ref{eq:lmd}) and (\ref{eq:lmb}), and assume instead
that the stellar $M/L$ ratio is constant for all the galaxies.
In this hypothesis, Eq. (\ref{eq:fpd2}) can be used to
predict how the contribution of the disk to the RC should scale with
surface brightness and scalelength.
Since we have, for an exponential disk: 
\begin{equation}
\protect\label{eq:vdisk}
V_d \sim [(M/L)\, I(0)\, R_d]^{0.5} \;\; ,
\end{equation}
and from Eq. (\ref{eq:fpd2}):
\[ V_{2} \sim I(0)^{0.41} R_d^{0.68} \;\; , \]
if $M/L$ is a constant, we derive 

\begin{equation}
\protect\label{eq:vsc}
\frac{V_d}{V_{2}} \sim \frac{I(0)^{0.09 \pm 0.07}}{R_d^{0.18 \pm 0.07}} \;\; , 
\end{equation}

\noindent
implying that the relative disk contribution to the RC should be the smallest
for large galaxies and, although with less significance, for low surface 
brightness ones\footnote{ 
From Eqs. (\ref{eq:fpd}) and (\ref{eq:fpd2}) we can see instead that, in
our RC fits, the ratio $V_d/V_2$ is about constant, i.e. independent of $R_d$
and $I(0)$: it amounts to about 85 \% on average, the typical ``maximum
disk'' value. As a consequence, the disk $M/L$ must increase with $R_d$.}. 
For the bulges, using Eqs. (\ref{eq:fpb}) and (\ref{eq:vdisk}), we find
a trend with size similar to the one expressed by Eq. (\ref{eq:vsc}).
Considering the range of 
parameters, rather uniformely covered by our sample (about three magnitudes 
in $I(0)$ and a factor of 4 in $R_d$), and assuming that the brightest and 
more compact disks contribute about 90\% to the overall RC at 2.2 $R_d$, then 
Eq. (\ref{eq:vsc}) implies that,
for the galaxies with the largest radii and faintest surface brightness, 
$V_d$ should be about 55\% of the observed rotation velocity.
In this hypothesis a ``maximum disk'' RC can be ruled out in most cases;
of course, to match the observed RC's, we have to postulate a conspicuous
increase with size of the relative amount of dark matter.
As already mentioned in Sect. \ref{kin}, the results by Bottema 
(\cite{botte}, \cite{botte2}) and by Courteau \& Rix (\cite{cou_rix}) 
support a scenario of 
this kind, with the disk contributing, on average, about 60\% of the 
rotation at 2.2 $R_d$'s. Assuming that this rule holds for our sample
as well, then, Eq. (\ref{eq:vsc}) predicts that the ratio 
$V_d/V_2$ should vary throughout the sample between about 0.45 and 0.75, 
in a well defined way according to the size and surface brightness of 
each disk. Neglecting the presence of the bulge, and assuming an infinitely
thin disk, the corresponding ratio of dark to visible mass 
within $R_2$ can be expressed as
\[\frac{M_h}{M_d} \simeq 0.8\left[\left( \frac{V_2}{V_d} \right)^{2} - 1 \right] \;\; .\]
This quantity changes by more than a factor of 8 in our sample, from 0.4 to 
3.3, whereas in the ``maximum disk'' hypothesis it stays roughly constant:
in principle, the predictions of a reliable model for galaxy formation could 
help to distinguish between two such different behaviours.

Of course, any intermediate scenario between the two extremes discussed
here (``maximum disk'' solutions and constant stellar $M/L$) could be 
considered as well. 
%We remind again that a systematic increase of the stellar 
%$M/L$ with luminosity is suggested, by 
%the colour--magnitude relation, and also from a comparison of the observed 
%RC's with the predictions of cosmological simulations (Navarro et al. 
%\cite{navarro}, Navarro \cite{nava}).
From the observational point of view, again, more detailed spectrophotometric 
data might better constrain
possible variations in the average stellar population of different
galaxies, and help to distinguish between the different possibilities.

% \begin{equation}
% \protect\label{eq:vscb}
% \frac{V_d}{V_b} \sim \frac{1}{I(0)^{0.13 \pm 0.12}R_d^{0.35 \pm 0.14}}
% \;\; ,
% \end{equation}

\section{Summary}

Using near--infrared images and rotation curves of a sample of 40 spiral 
galaxies, we have determined the scaling relations, between structural 
and kinematic parameters of bulges and disks, analogous to the Fundamental 
Plane of elliptical galaxies.
The accuracy of the  disk FP as a distance indicator, for this set of data and
our photometric decompositions, is comparable but slightly lower than the one 
attained by the Tully--Fisher relation. This suggests that the fine tuning 
between dark and visible components at the basis of the various scaling 
relations is more effective for global parameters.
Also, we deduce that (a) either the stellar mass-to-light ratio of the disk
increases with $R_d$, or (b) the disk contribution to the observed RC 
decreases according to Eq. (\ref{eq:vsc}) for galaxies of large size.
A similar behaviour is observed for the bulges.

\begin{acknowledgements}
We would like to thank the referee, A. Bosma, for his careful reading of 
the manuscript, C. Giovanardi and L. Hunt for insightful comments and 
suggestions, and Stephane Courteau for having provided his data
in electronic format.
This research was partially funded by ASI Grant ARS--98--116/22.
Partial support during residency of G.M. at Cornell University was obtained via 
the NSF grant AST96-17069 to R. Giovanelli.
\end{acknowledgements}

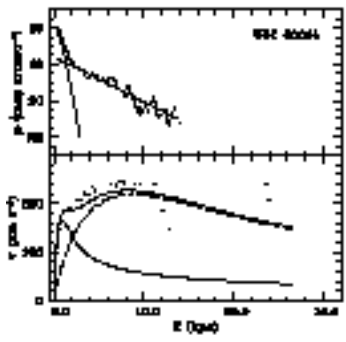
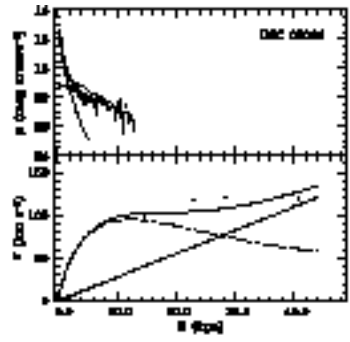
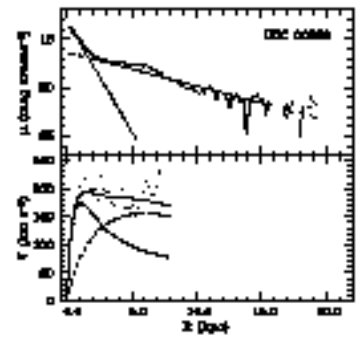
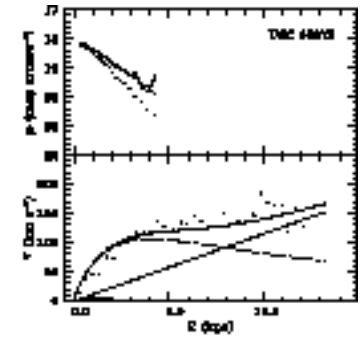
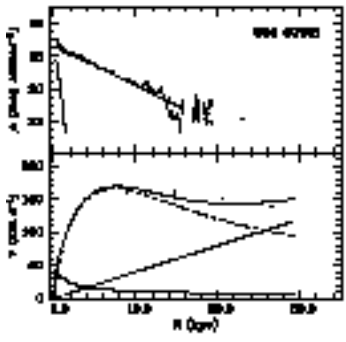
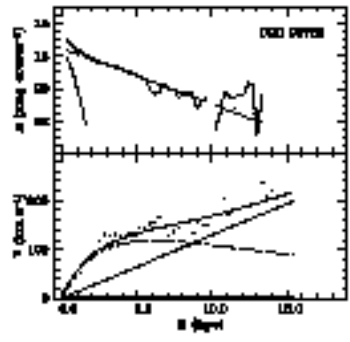
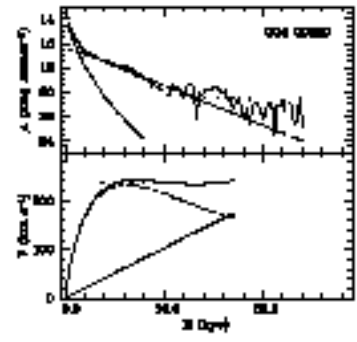
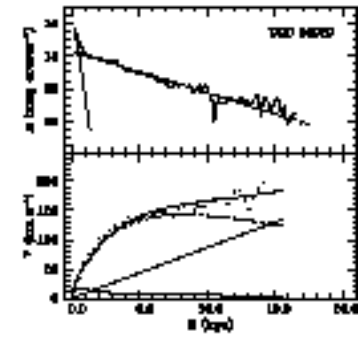
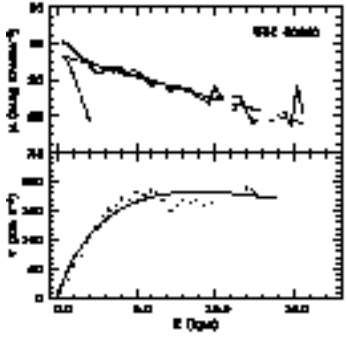
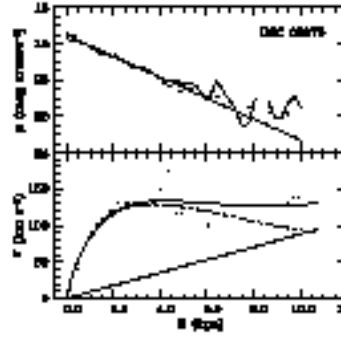
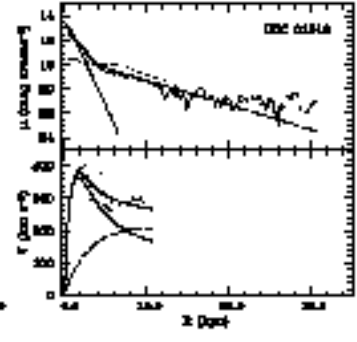
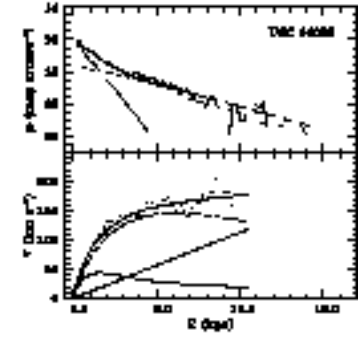
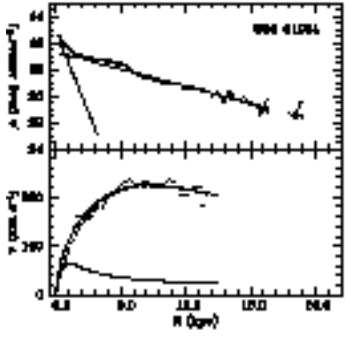
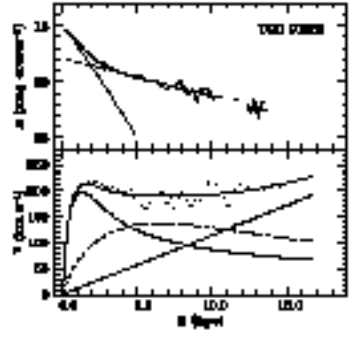
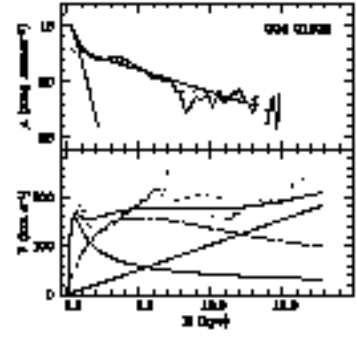
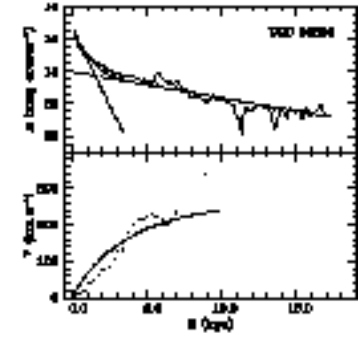
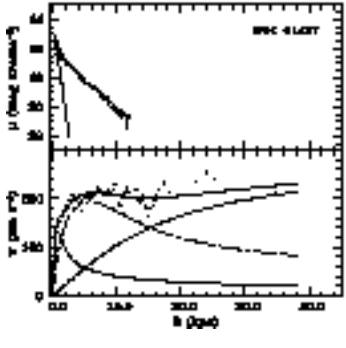
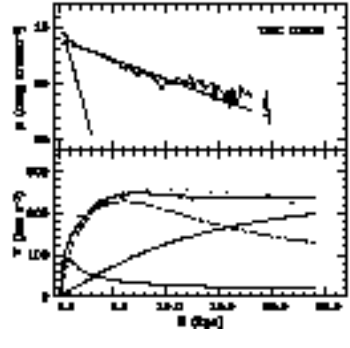
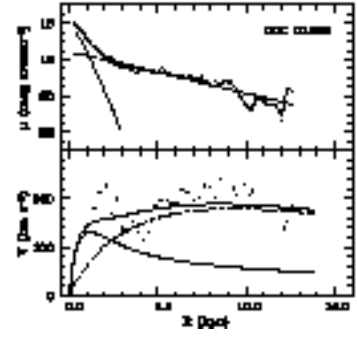
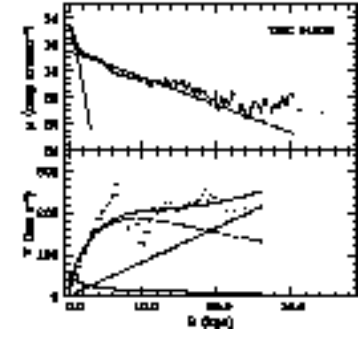
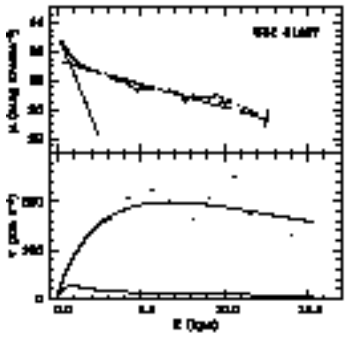
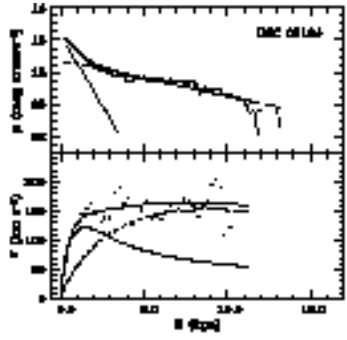
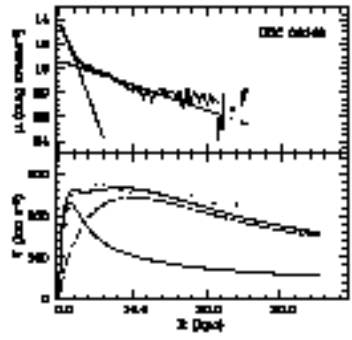
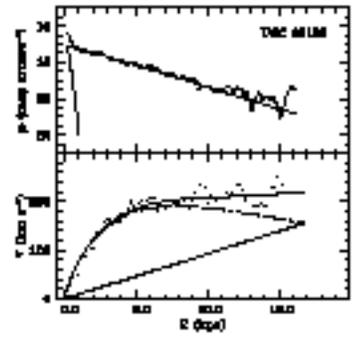
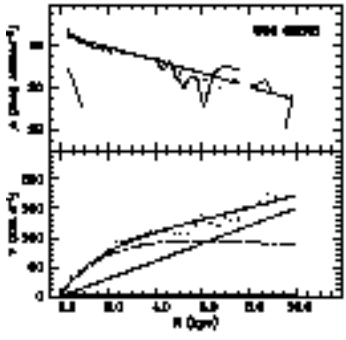
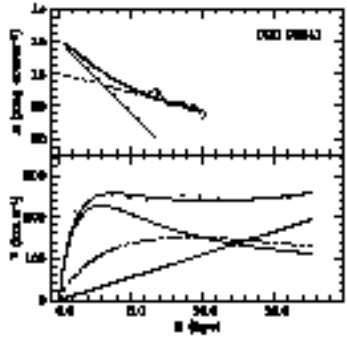
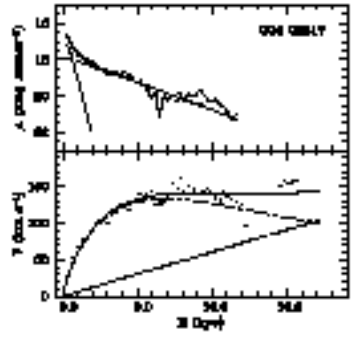
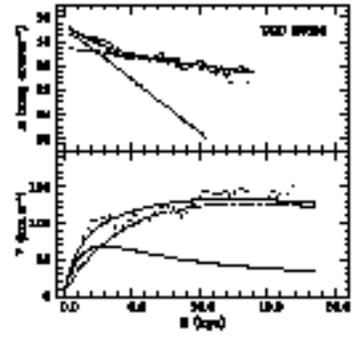
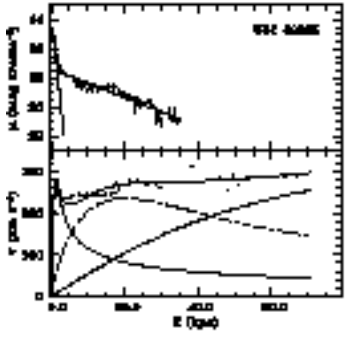
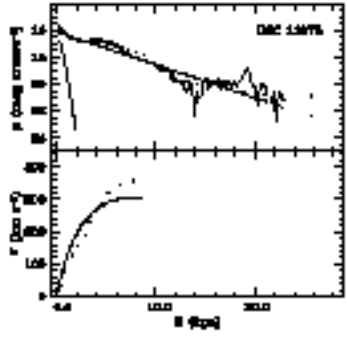
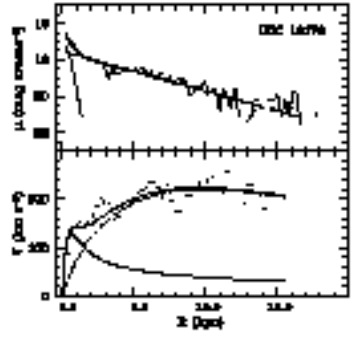
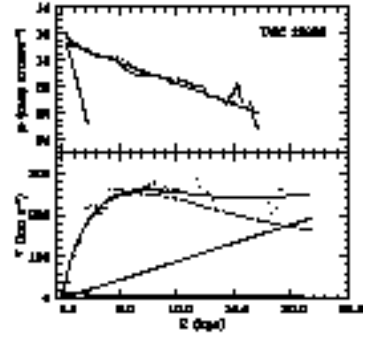
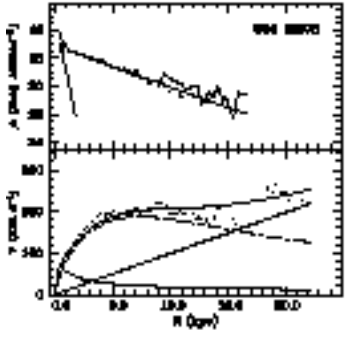
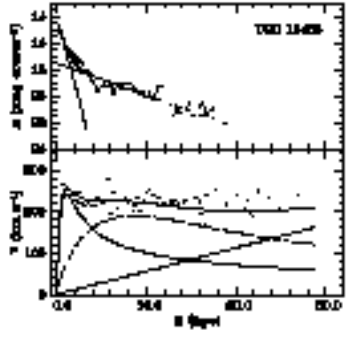
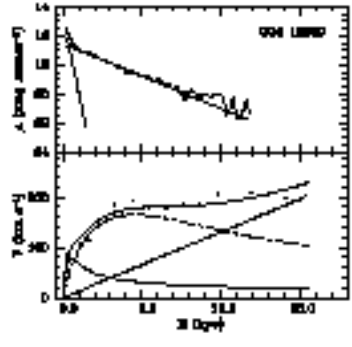
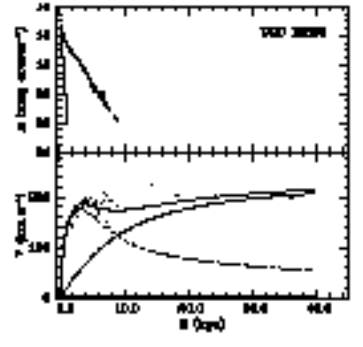
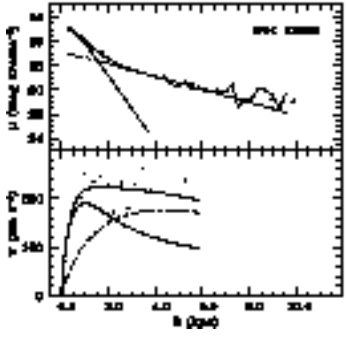
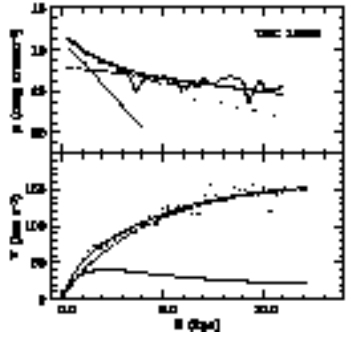
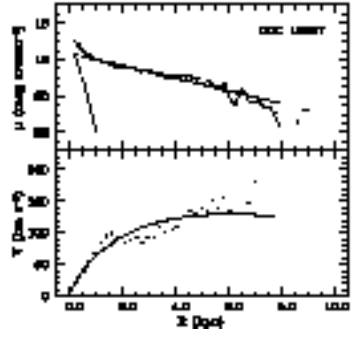
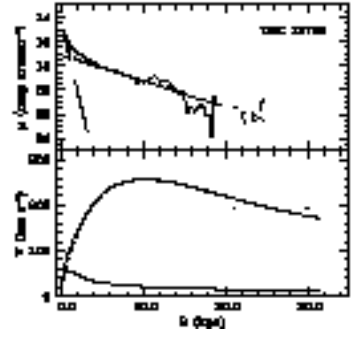
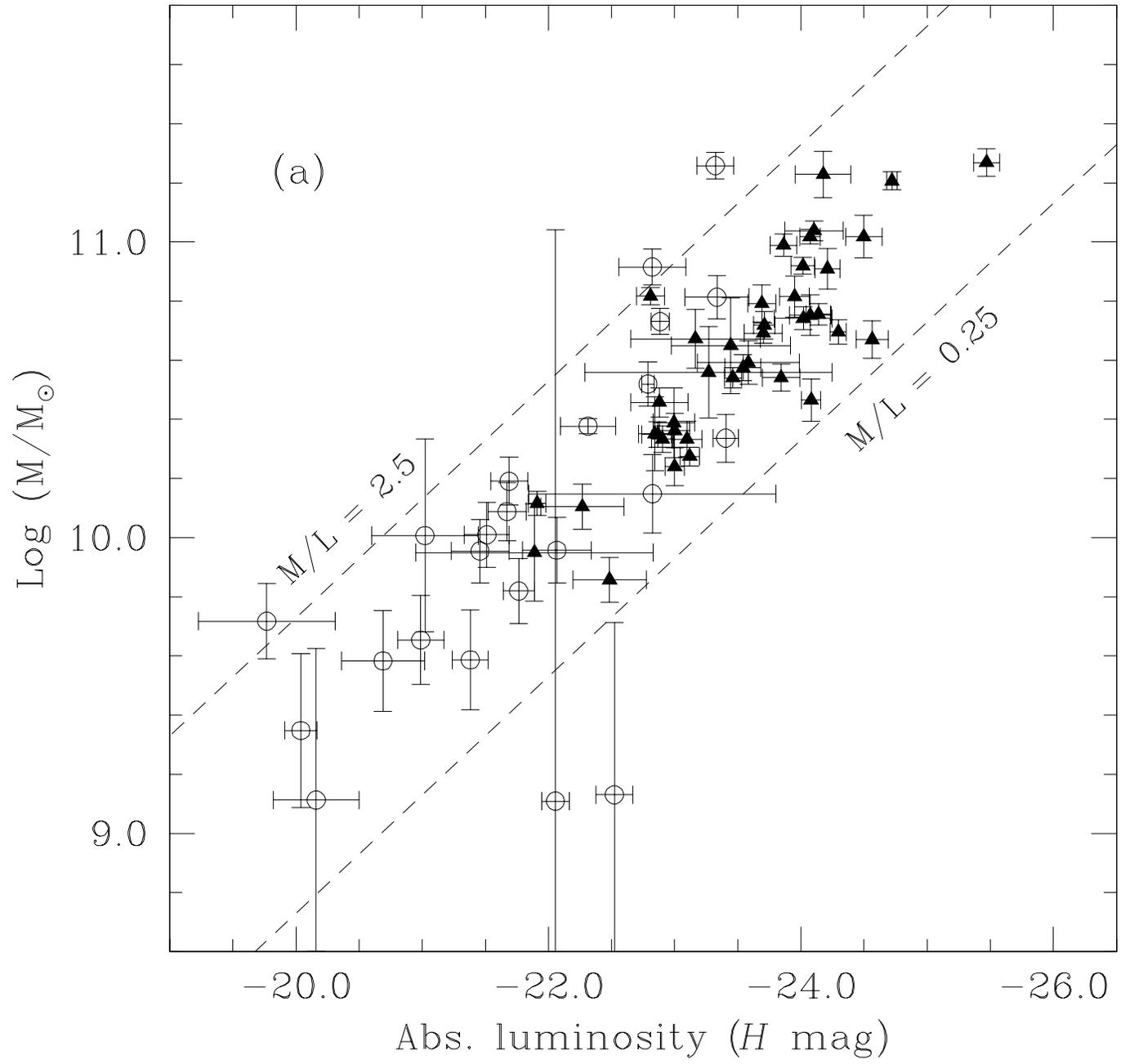
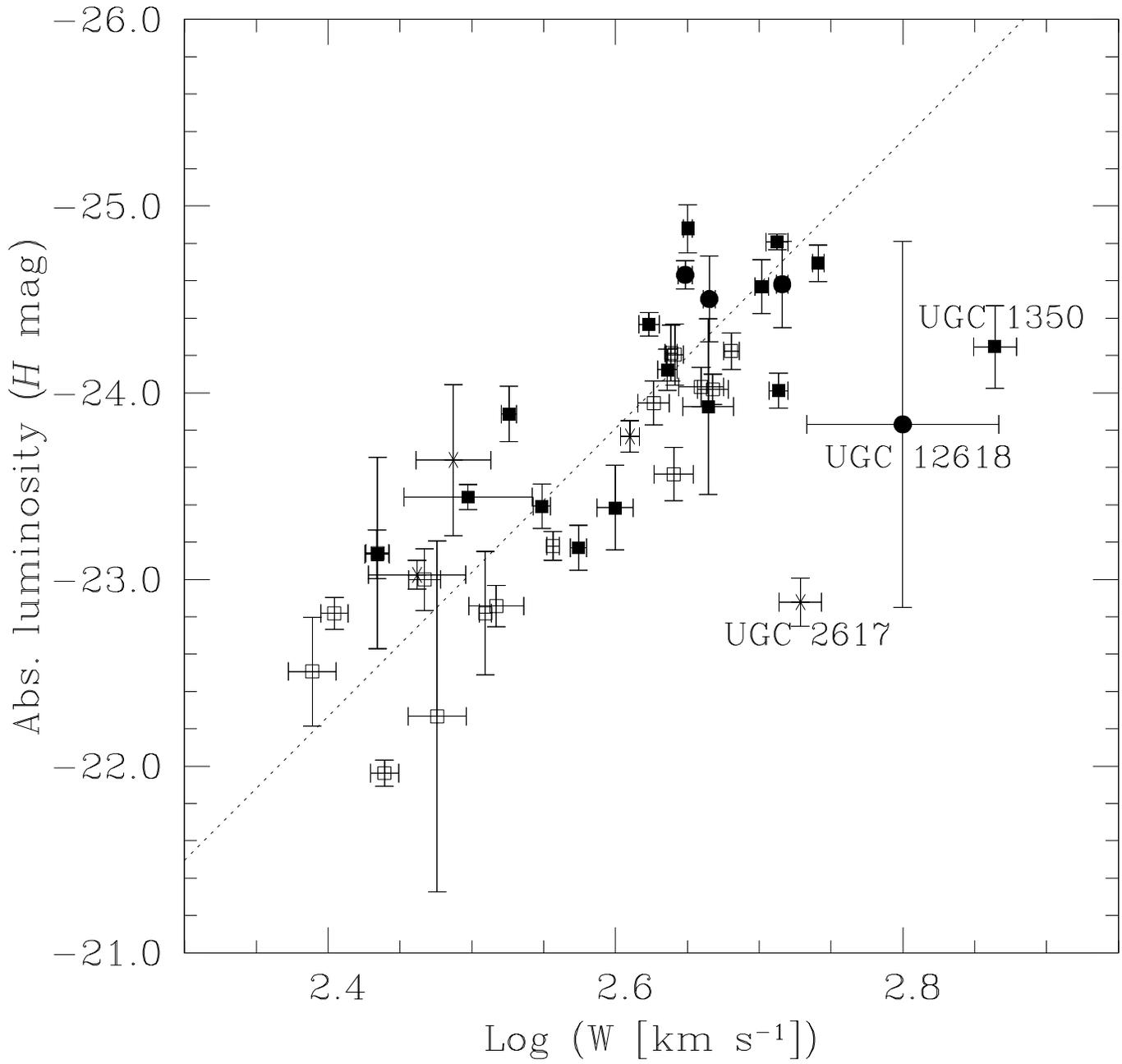
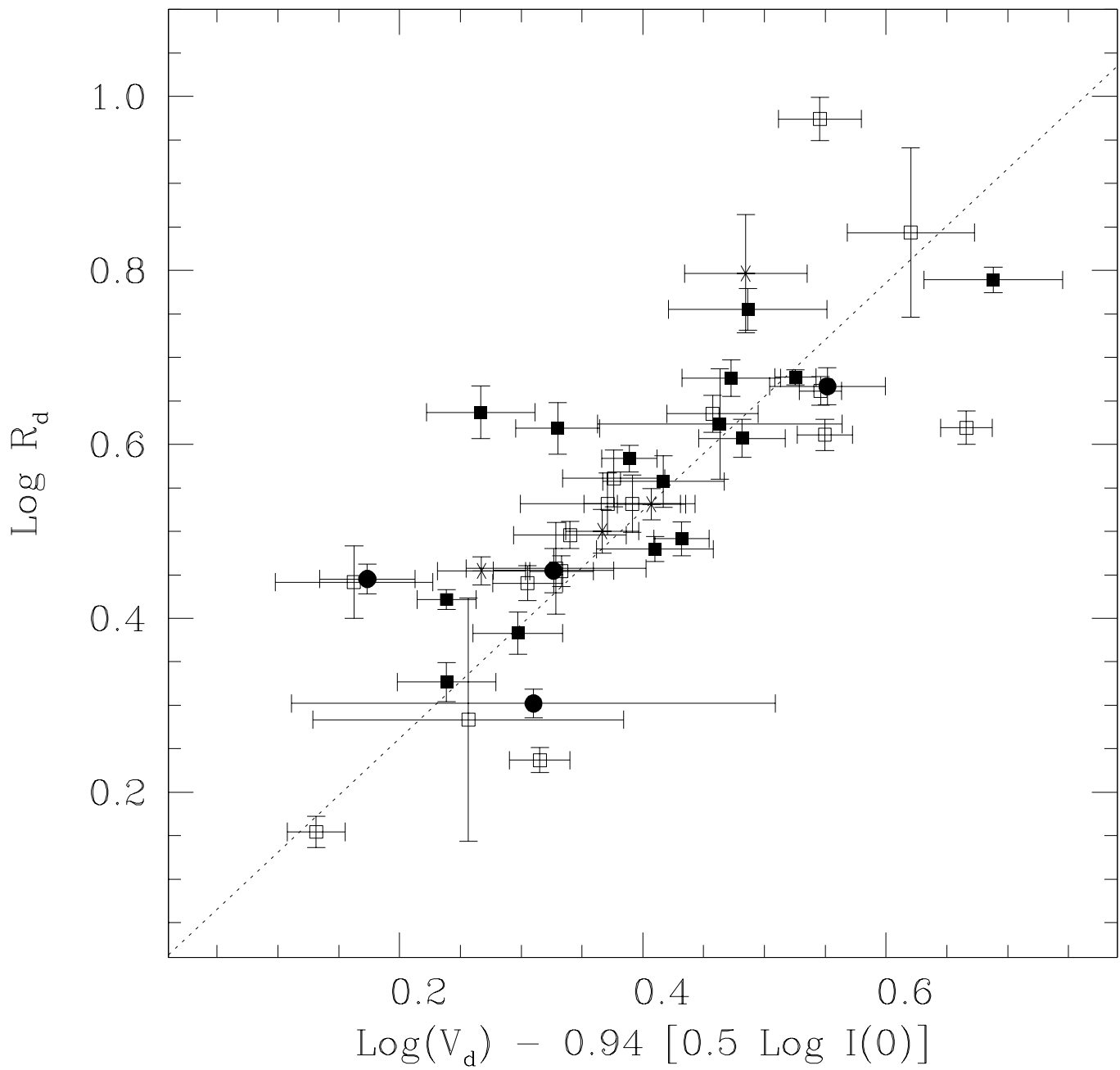
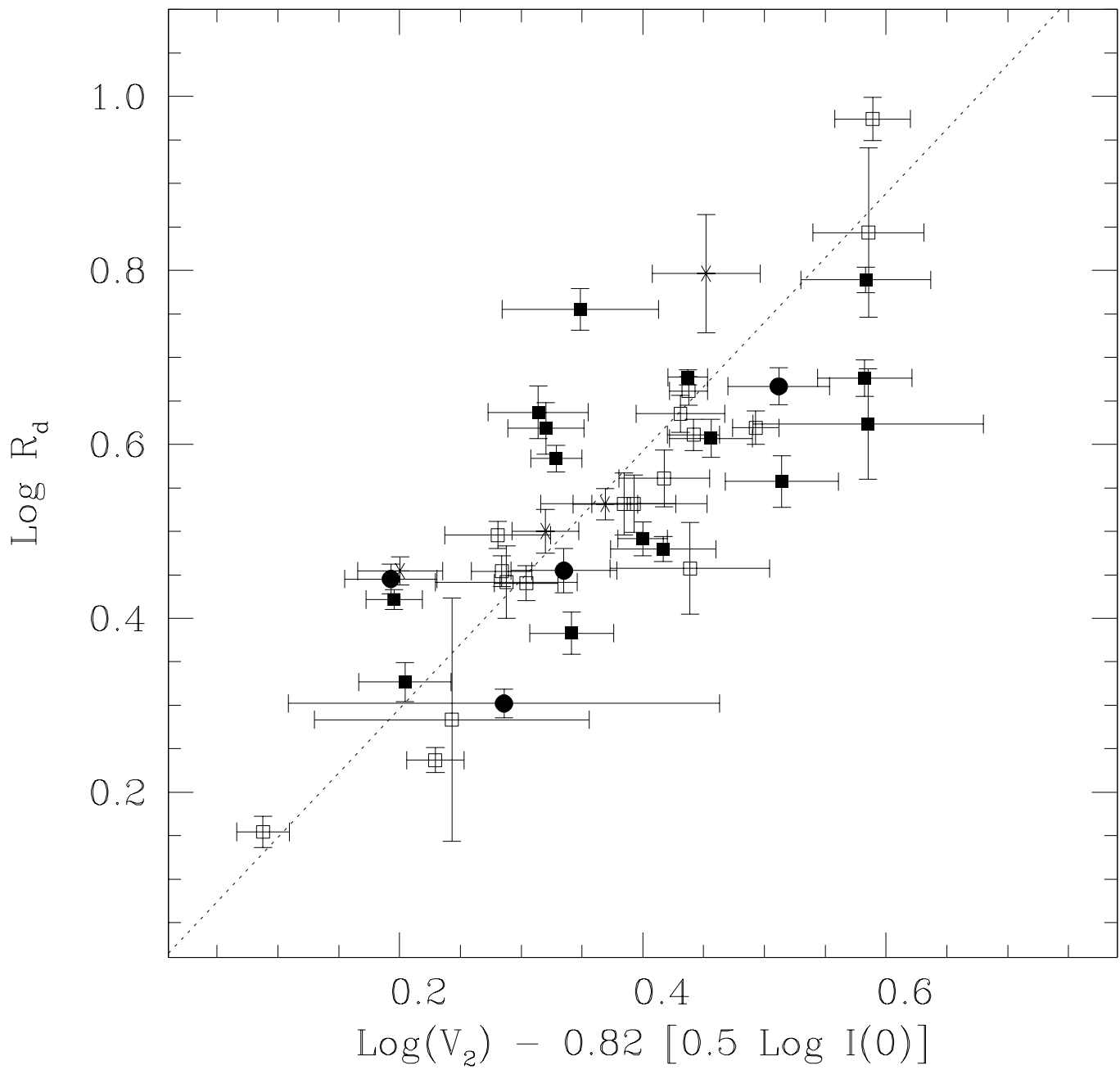
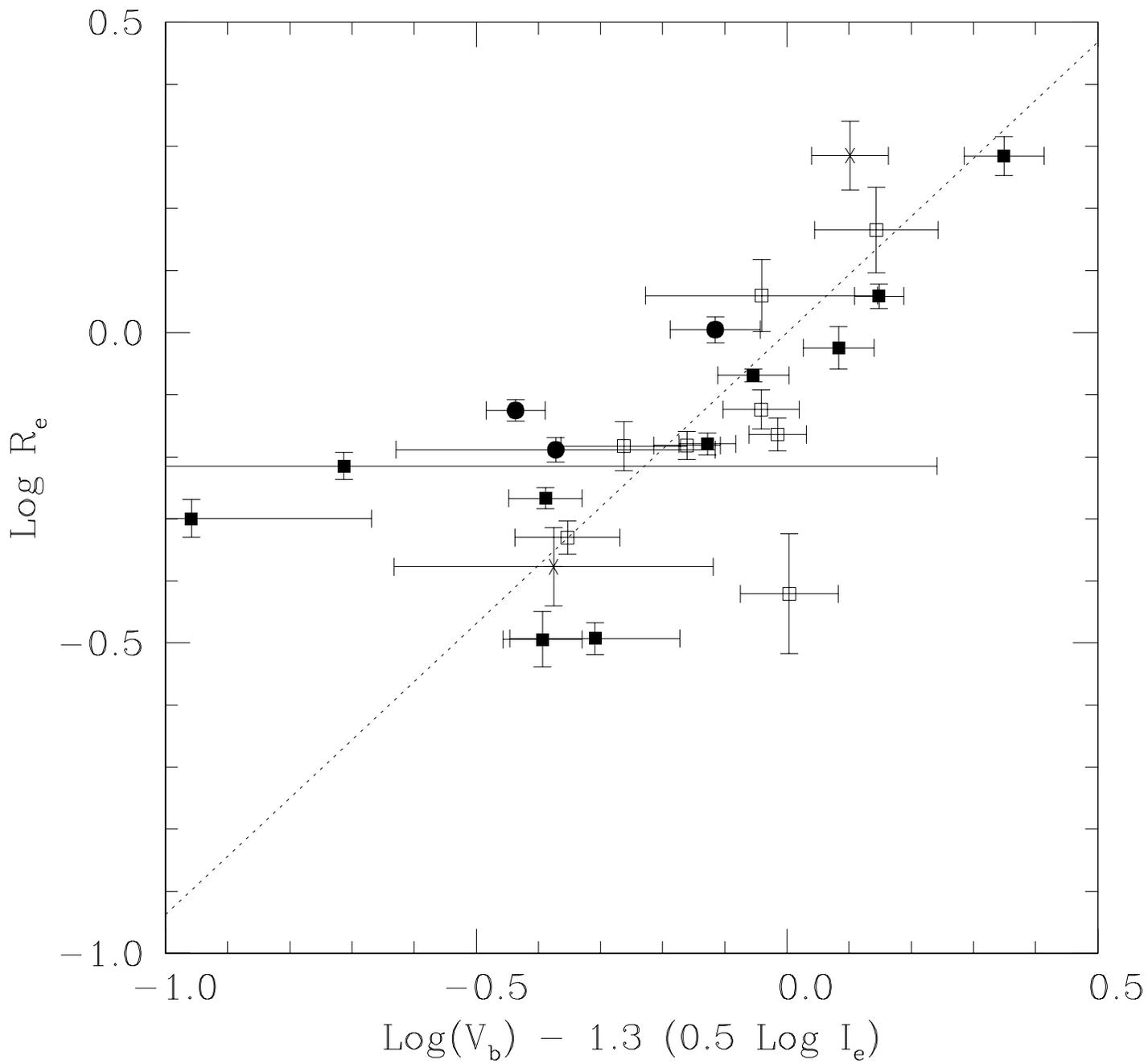
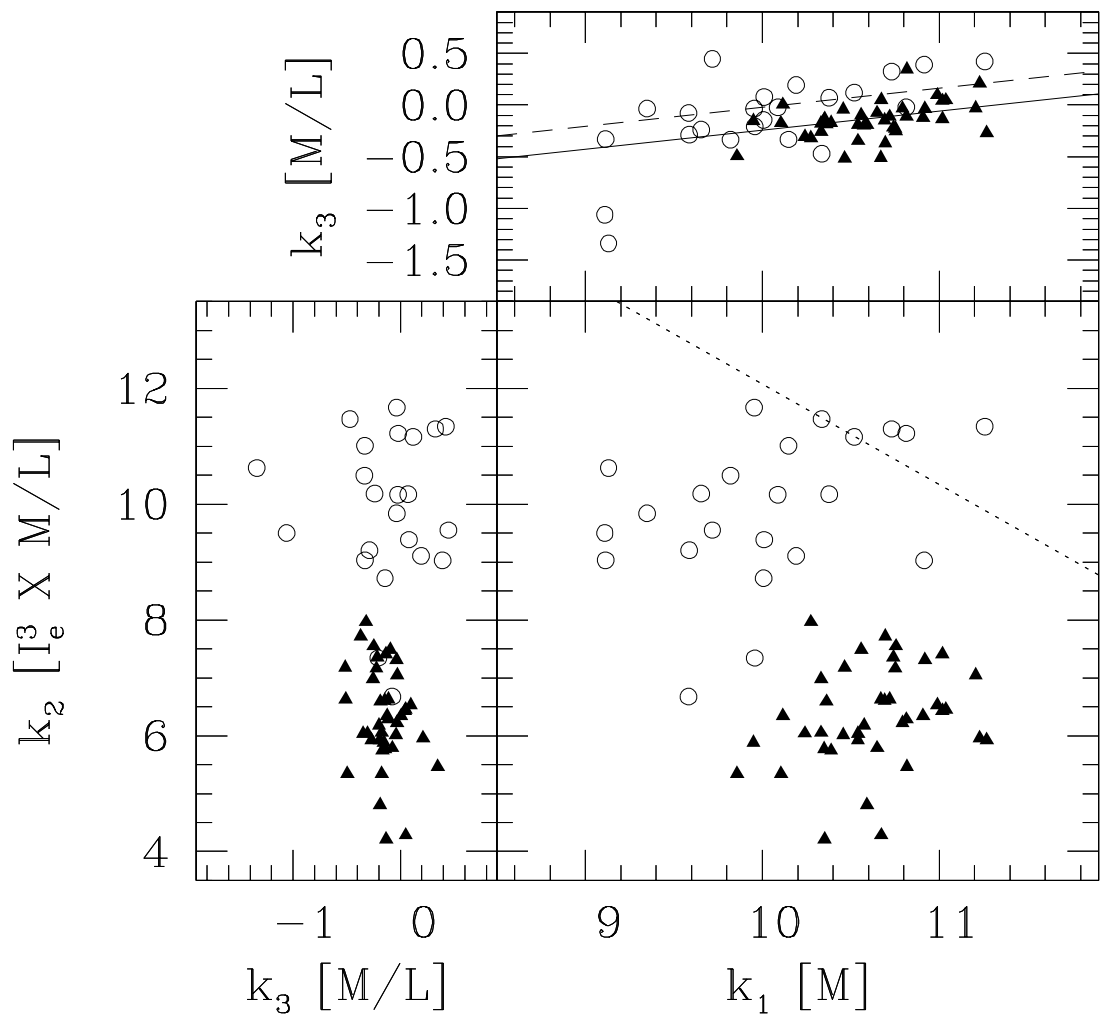
\end{document}